%% file: main.tex
\documentclass{article}

\usepackage{arxiv}

\usepackage[utf8]{inputenc} 
\usepackage[T1]{fontenc}    
\usepackage{lmodern}
\usepackage{hyperref}       
\usepackage{url}            
\usepackage{longtable}
\usepackage{booktabs}       
\usepackage{multirow}
\usepackage{caption}

\usepackage{amsfonts,amsmath,amssymb}
\usepackage{nicefrac}       
\usepackage{microtype}      
\usepackage{scalerel}
\usepackage{color}

\usepackage{natbib}
\usepackage{pgf}
\usepackage{tikz}
\usetikzlibrary{arrows,automata}
\usetikzlibrary{positioning}

\tikzset{
    state/.style={
           rectangle,
           rounded corners,
           draw=black, very thick,
           minimum height=2em,
           inner sep=2pt,
           text centered,
           }
}

\title{A review of Bayesian perspectives on sample size derivation for confirmatory trials}

\author{
  Kevin Kunzmann \thanks{
        MRC Biostatistics Unit, University of Cambridge,
        Cambridge Institute of Public Health,
        Forvie Site, Robinson Way,
        Cambridge Biomedical Campus,
        Cambridge CB2 0SR,
        United Kingdom
    }\\
    MRC Biostatistics Unit \\
    University of Cambridge \\
    \texttt{kevin.kunzmann@mrc-bsu.cam.ac.uk} \\
    \And
    Michael J. Grayling \\
    Population Health Sciences Institute \\
    Newcastle University \\
    \texttt{michael.grayling@newcastle.ac.uk} \\
    \And
    Kim May Lee \\
    Pragmatic Clinical Trials Unit\\
    Queen Mary University of London\\
    \texttt{k.m.lee@qmul.ac.uk} \\
    \And
    David S.\ Robertson \\
    MRC Biostatisics Unit\\
    University of Cambridge\\
    \texttt{david.robertson@mrc-bsu.cam.ac.uk} \\
    \And
    Kaspar Rufibach \\
    Methods, Collaboration, and Outreach Group (MCO) \\
    Department of Biostatistics\\
    F. Hoffmann-La Roche, Basel\\
    \texttt{kaspar.rufibach@roche.com}
     \And
    James M. S. Wason \\
    Population Health Sciences Institute \\
    Newcastle University\\
    and \\
     MRC Biostatistics Unit \\
    University of Cambridge \\
    \texttt{james.wason@newcastle.ac.uk}
}

\renewcommand{\Pr}{\operatorname{Pr}}
\newcommand{\E}{\operatorname{E}}
\newcommand{\st}{\operatorname{subject\ to:\ }}
\newcommand{\argmin}[1]{\operatornamewithlimits{argmin}_{#1}}
\newcommand{\argmax}[1]{\operatornamewithlimits{argmax}_{#1}}

\newcommand{\EP}{\operatorname{EP}}

\begin{document}

\maketitle

\captionsetup{width=\textwidth}

\begin{abstract}
Sample size derivation is a crucial element of the planning phase of any confirmatory trial.
A sample size is typically derived based on
constraints on the maximal acceptable type~I error rate and a
minimal desired power.
Here, power depends on the unknown true effect size.
In practice, power is typically calculated either for the smallest relevant effect size or a likely point alternative.
The former might be problematic if the minimal relevant effect is close to the null, thus requiring an excessively large sample size.
The latter is dubious since it does not account for the \textit{a~priori} uncertainty about the likely alternative effect size.
A Bayesian perspective on the sample size derivation for a frequentist
trial naturally emerges as a way of reconciling arguments about the relative \textit{a~priori} plausibility of alternative effect sizes with ideas based on the relevance of effect sizes.
Many suggestions as to how such `hybrid' approaches could be implemented in practice have been put forward in the literature.
However, key quantities such as assurance, probability of success, or expected power are often defined in subtly different ways in the literature.
Starting from the traditional and entirely frequentist approach to sample size derivation, we derive consistent definitions for the most commonly used
`hybrid' quantities and highlight connections, before discussing and demonstrating their use in the context of sample size derivation for clinical trials.
\end{abstract}


\section{Introduction}
\label{sec:intro}

Randomised Controlled Trials (RCTs) are the gold-standard study design for evaluating the effectiveness and safety of new interventions.
Despite their successful history of demonstrating the benefits
and uncovering the risks of new treatments,
they face substantial challenges.
Recent evidence has shown that the success rate of RCTs is low~\citep{wong2019estimation}.
This low success rate and the increasing cost of conducting RCTs is resulting in large real-term increases in the cost of drug development~\citep{dimasi2016innovation}.
Since the sample size of a trial is a key determinant of the chances of detecting a treatment effect (if it is present) and
an important cost factor,
the choice of an adequate sample size enables more
economic drug development.
Purely economic arguments would suggest a utility based approach
as discussed in~\citet{lindley-1997}, for example.
However, specifying utility functions, particularly in investigator-sponsored studies concerned with non-drug interventions,
can be hard in practice.
These practical problems, the desire to maintain minimal ethical standards, as well as recommendations of health authority guidelines result in the vast majority of confirmatory clinical trials deriving their sample size based on desired conditional type~I and type~II error rates.
For instance, a randomised clinical trial with an unnecessarily large sample size (`overpowered') is unethical if the treatment shows a substantial effect and the consequences of being randomised to the control arm are severe.
Thus the issue of selecting an appropriate sample size is a vitally important part of conducting a clinical trial.

The traditional approach to determining the required sample size for a trial is to choose a point alternative and
derive a sample size such that the probability to reject the null hypothesis exceeds a certain threshold (typically 80\% or 90\%) while maintaining a certain maximal type~I error rate (typically 2.5\% one-sided).
The maximal type~I error rate is usually realised at the boundary of the null hypothesis and thus available immediately.
The type~II error rate, however, critically depends on the choice of the (point) alternative.
There are at least two ways of justifying the choice of point alternative.
The first is based on a relevance argument, which requires the specification of a minimal clinically relevant difference (MCID).
Since the probability to reject the null hypothesis is typically monotonic in the effect size, determining the sample size such that the desired probability to reject the null is exceeded at the MCID implies that the power for all other relevant differences will be even larger.
Guidance has recently been published on choosing the MCID~\citep{cook2018delta2}, but making this choice may still be difficult in practice.

The second perspective is based on \textit{a~priori} considerations about the likelihood of the treatment effect.
Here, an \textit{a~priori} likely effect size is used as the point alternative.
This implies that the resulting sample size might be too small to detect smaller but still relevant differences reliably.
On the other hand, the potential savings in terms of sample size might outweigh the risk of ending up with an underpowered study.
The core difference between these approaches is that basing the required sample size on the MCID might be ineffective if prior evidence for a larger effect is available, but the MCID is chosen based on \emph{relevance} arguments and is thus not subject to uncertainty.
In contrast, choosing the point alternative based on considerations about the relative \textit{a~priori} likelihood of effect sizes implies that there is an inherent uncertainty about the effect size that can be expected -- otherwise no trial would be needed in the first place.

Depending on the study objective, the sample size can also be
derived based on entirely different considerations.
For example, studies that aim to establish new diagnostic tools or biomarkers would rather target a certain width of the confidence interval for the AUC~\citep{obuchowski1998}.
Similarly, studies aimed at estimating population parameters with sufficient precision should target the standard error of the estimate, rather than deriving a sample size based on power arguments~\citep{Thompson2012,grouin-2007}.
These approaches are beyond the scope of this manuscript and we will only discuss sample size derivation based on error rate considerations.

To make things more tangible, consider the case of a simple
one-stage, one-arm $Z$-test.
Let $X_i$, $i=1,\ldots,n$, be $iid$ observations with mean $\theta$ and known standard deviation $\sigma$.
Under suitable regularity conditions, their mean is asymptotically normal and
{\small $Z_n := \sqrt{n}\,\big(\overline{X} - \theta_0\big)/\sigma\stackrel{\cdot}{\sim}\mathcal{N}\big(\theta_n,1\big)$} where $\theta_n := \sqrt{n}(\theta - \theta_0)/\sigma$.
Assume that interest lies in testing the null hypothesis $\mathcal{H}_0:\theta \leq \theta_0 = 0$ at a one-sided significance level of $\alpha$.
\newcommand{\criticalvalue}{z_{1-\alpha}}
The critical value for rejecting $\mathcal{H}_0$ is given by the $(1-\alpha)$-quantile of the standard normal distribution, $\criticalvalue$, and is independent of $n$.
\newcommand{\mcid}{\theta_{\scaleto{\operatorname{MCID}}{3.5pt}}}
The probability of rejecting the null hypothesis for given $n$ and $\theta$ is
\begin{align}
\label{eq:prob-to-reject}
    \Pr_{\theta}[\,Z_n > \criticalvalue\,] = 1 - \Phi\big(\criticalvalue - \theta_n\big) = \Phi\big(\theta_n - \criticalvalue\big),
\end{align}
where $\Phi$ is the cumulative distribution function (CDF) of the standard normal distribution.
Often, $\Pr_{\theta}[\,Z_n > \criticalvalue\,]$ is seen as a function of $\theta$ and termed the `power function'.
This terminology may lead to confusion when considering parameter  values $\theta\leq\theta_0$ and $\theta\geq\mcid$, since the probability
to reject the null hypothesis corresponds to the type~I error rate in the former case and classical `power' in the latter.
For the sake of clarity we will thus use the neutral term `probability to reject'.

\newcommand{\altv}{\theta_{\scaleto{\operatorname{alt}}{3.5pt}}}
Assume that a point alternative $\altv>\theta_0$ is given.
A sample size can then be chosen as the smallest sample size that results in a probability to reject of at least $1-\beta$ at $\altv$
\begin{align}
    n^*_{\altv} :=\quad &\argmin{n}\ &&n \nonumber\\
          &\st         &&\Pr_{\altv}[\,Z_n > \criticalvalue\,] \geq 1 - \beta \ .
          \label{eq:simle-pwr-cnstr}
\end{align}
Since $\Pr_{\theta}[\,Z_n > \criticalvalue\,]$ is monotone in $\theta$, the probability to reject the null hypothesis for $\theta>\altv$ is also greater than $1-\beta$.
Thus, if $\altv=\mcid$, the null hypothesis can be rejected for all clinically relevant effect sizes with a probability of at least $1-\beta$.
This approach requires no assumptions about the \textit{a priori} likelihood of the value of $\theta$ but only
about $\mcid$ and the desired minimal power level
(see also~\cite[Section 3]{chuang2011} and~\citealp{chuang-2006}).
However, the required sample size increases quickly as $\mcid$ approaches $\theta_0$.
In almost all practically relevant situations, a maximal feasible sample size is given (e.g., due to budget constraints), which might not be sufficient to achieve the desired power if $\mcid$ is close to $\theta_0$.
This problem might arise when considering overall survival in oncology trials, for example.
However, it may then be hard to justify a value for $\mcid$ much larger than $\theta_0$ since almost any improvement in overall survival can be considered relevant
\footnote{It is important to distinguish between the (unknown) true effect size and the observed effect size.
It might still be reasonable to additionally require a certain deviation of the observed effect from the null, e.g.~a hazard ratio of less than 0.85.}.
The problem becomes even more pressing if the null hypothesis is defined as $\mathcal{H}_0':\theta \leq \mcid$, i.e.,
if the primary study objective is to demonstrate a clinically important effect.
In either case it is clearly impossible to derive a feasible sample size based on the minimal clinically important difference alone~\citep{chuang2011}.
Formulating a principled approach to eliciting a sample size in situations like these is difficult, and in practice trialists may resort to back-calculating an effect size in order to achieve the desired power given the maximum feasible sample size~\citep{lenth2001,lan-2012,grouin-2007}.
One way of justifying a smaller sample size is to consider a point alternative $\altv>\mcid$ based on \textit{a~priori} likelihood arguments: if there is prior evidence for effect sizes larger than $\mcid$, determining the sample size based on $\mcid$ might well be inefficient and lead to an unnecessarily large trial.
Therefore, planning of the required sample size is often based on a single point alternative $\altv$, $\altv > \mcid$.
Yet, this pragmatic approach is unsatisfactory in that it ignores any uncertainty about the effect
size~\citep{lenth2001}.

In the following, we review approaches to sample size derivation that do account for \textit{a~priori} uncertainty via a prior density for the effect size.
We propose a framework encompassing the most relevant
quantities discussed in this context, give precise definitions of the terms, and highlight connections between individual items.
Where necessary, we refine existing definitions to improve overall consistency.
Note that we exclusively focus on what is usually termed
a `hybrid' Bayesian-frequentist approach
to sample size derivation~\citep{spiegelhalter-2004}.
This means that, although, Bayesian arguments are used to derive
a sample size under uncertainty about the true effect size,
the final analysis is strictly frequentist.
After introducing the individual concepts in detail,
a structured overview of all quantities considered is provided in Figure~\ref{fig:diagram}.
We then present a review of the literature on the subject, showcasing the confusing diversity of terminology used in the field and relating our definitions back to the existing literature.
Finally, we present some numerical examples and conclude with a discussion.

\section{A Bayesian argument for the choice of the point alternative}
\label{sec:quantile-approach}

\newcommand{\rpr}{\operatorname{RPR}}

One way of incorporating planning uncertainty is to make assumptions about the relative \textit{a~priori} likelihood of the unknown effect size.
This approach can be formalised within a Bayesian framework by seeing the true effect size $\theta$ as the realisation of a random
variable $\Theta$ with prior density $\varphi(\theta)$.
This means that the CDF of
$\Theta$ is given by $\Pr_{}[\,\Theta\leq x\,] = \int_0^x \varphi(\theta) \operatorname{d} \theta$.
At the planning stage, the probability to reject the null hypothesis is then given by the random variable $\rpr(n)$, the `random probability to reject':
\begin{align}
    \rpr(n):=\Pr_{\Theta}[\,Z_n > \criticalvalue\,]\ .
\end{align}
We explicitly denote this quantity as `random' to emphasise the distinction between the
(conditional on $\Theta=\theta$) probability to reject given in equation~\eqref{eq:prob-to-reject} and the
unconditional `random' probability to reject.
The variation of the random variable $\rpr(n)$ reflects the \textit{a~priori} uncertainty about the unknown underlying effect size that is encoded in the prior density $\varphi(\cdot)$ of the random variable $\Theta$.
We prefer the term `random probability to reject' over `random power' since $\rpr(n)$ is unconditional on the effect size, and consequently does not distinguish between rejections under the null hypothesis and under relevant effect sizes.
\newcommand{\rpower}[1]{\operatorname{RPow}({ #1 })}
Instead, we define the conditional random variable `random power' as
\begin{align}
    \rpower{n}:= &\ \Pr_{\, \Theta\geq\mcid\, }[\,Z_n > \criticalvalue\,] \\ =& \ \rpr(n) \mid \Theta\geq\mcid.
\end{align}
This definition more closely resembles the concept of frequentist power since it conditions on a relevant effect size.

Determining the required sample size based on a point alternative as outlined in the introduction evaluates the probability to reject the null hypothesis solely on $\altv$.
This can be understood as conditioning the random probability to reject, or the random power, on $\Theta=\altv$, i.e., to consider $\big(\rpr(n) \mid \Theta = \altv \big) = \big(\rpower{n} \mid \Theta = \altv \big)$.
Due to conditioning on a single parameter value $\altv\geq\mcid$, under any prior density with $\varphi(\altv)>0$, both random variables (almost surely) reduce to the deterministic expression $\Pr_{\altv}[\,Z_n > \criticalvalue\,]$ which is often termed `power' in a frequentist context.
Basing the sample size derivation on this quantity means that the probability to reject the null hypothesis for relevant values of $\theta\geq\mcid, \theta\neq\altv$ is completely ignored and the \textit{a~priori} evidence encoded in $\varphi(\cdot)$ is not used.

\citet{spiegelhalter1986} have pointed out that a power constraint for sample size derivation could be computed based on \textit{``[...] a somewhat arbitrarily chosen location parameter of the [prior] distribution (for example the mean, the median or the 70th percentile).''}
Using a location parameter of the unconditional prior distribution of $\Theta$, however, might lead to situations where no sample size can be determined if the location parameter lies within the null hypothesis.
Here, we follow a similar idea but motivate the choice of location parameter in terms of the \textit{a~priori} distribution of random power.
To this end, let
\newcommand{\quant}[1]{{\operatorname{Q}_{ #1 }}}
\begin{align}
    \quant{ 1-\gamma }[\ \rpower{n} \,] \ := \
    \inf_{x}\ \Pr_{\varphi(\cdot)}\big[\, \rpower{n} \geq x \,\big] \geq \gamma\
\end{align}
be the $(1-\gamma)$-quantile of the random power ($\rpower{n}$).
Furthermore, let
\begin{align}
    \varphi(\theta\,|\,\Theta\geq \mcid):= \frac{\displaystyle \boldsymbol{1}_{\theta\geq \mcid} \ \varphi(\theta)}{\displaystyle \int_{\mcid}^\infty \varphi(x)\ \operatorname{d} x } \label{eq:conditional-prior-density}
\end{align}
be the conditional prior density of $\big(\Theta \mid \Theta \geq \mcid \big)$.
We choose to make the dependency of \begin{align}
    \Pr_{\varphi(\cdot)}\big[\, \rpower{n} \geq x \,\big] = \int_{-\infty}^{\infty} \varphi(\theta\,|\,\Theta\geq \mcid)\, \Pr_{ \theta}[\,Z_n > \criticalvalue\,] \operatorname{d}\theta
\end{align}
on the prior density explicit by using the index `$\varphi(\cdot)$' since the random parameter $\Theta$ does not appear directly in the description of the event `$\rpower{n} \geq x$'.
Whenever $\Theta$ appears explicitly, we omit the index since the dependency on~$\varphi(\cdot)$ is then clear from the context.
The expression ${\Pr_{\varphi(\cdot)}\big[\, \rpower{n} \geq x \,\big]}$ is a real number in ${[\,0, 1\,]}$ and thus different from  $\rpr(n)=\Pr_{\Theta}[\,Z_n > \criticalvalue\,]$ which is a random variable.

If a sample size was then chosen such that $\quant{1-\gamma}[\,\rpower{n}\,]\geq 1 - \beta$, the  \textit{a~priori} probability of exceeding a probability to reject of $1-\beta$ given a relevant effect would be, by definition, at least $\gamma$.
The required sample size for this approach is the solution of
\begin{align}
    \argmin{n}\quad &n \nonumber\\
    \st \quad &\quant{1-\gamma}[\,\rpower{n}\,] \geq 1 - \beta \ .
\end{align}
Since $\Pr_{\theta}[\,Z_n > \criticalvalue\,]$ is monotonic in $\theta$, this problem is equivalent to solving
\begin{align}
    n_{\gamma}^* \quad := \quad & \argmin{n} \quad &&n \nonumber\\
    &\st \quad &&\Pr_{\ \quant{1 - \gamma}[\, \Theta \geq \mcid \,]}\big[\,Z_n > \criticalvalue\,\big] \geq 1 - \beta \ ,
\end{align}
where $\quant{1 - \gamma}[\, \Theta \geq \mcid \,]$ is the $(1-\gamma)$-quantile of the prior distribution of the random variable $\Theta$ conditional on a relevant effect.
Consequently, this `prior quantile approach' can be used with any existing frequentist sample size derivation formula.
It~is merely a formal Bayesian justification for determining the sample size of a trial based on a point alternative ${\altv:=\quant{1 - \gamma}[\, \Theta \geq \mcid \,]\geq\mcid}$.
The prior quantile approach reduces to powering on $\mcid$ whenever the target power needs to be met with absolute certainty for all relevant effect sizes ($\gamma=1$).
One may thus see the prior quantile approach as a principled relaxation of powering on $\mcid$.

The approach differs from Spiegelhalter and Freedman's suggestions in two key aspects.
Firstly, the point alternative naturally emerges as a quantile of the prior \emph{conditional on a relevant effect} by imposing a lower boundary on the \textit{a~priori} probability to undershoot the target power.
This intuitively makes sense since a large probability to reject is only desirable when the underlying $\theta$ is relevant.
This also ensures that $\quant{1 - \gamma}[\, \Theta \geq \mcid \,]>\mcid\geq\theta_0$ for $\gamma<1$ and thus guarantees a finite sample size irrespective of the choice of prior.
Secondly, to guarantee a more than 50\% chance of exceeding the target power, the conditional prior quantile will typically be chosen smaller (i.e.,~$\gamma>0.5$) than the conditional median ($\gamma=0.5$) which was discussed by Spiegelhalter and Freedman.

\section{Probability of success and expected power}
\label{sec:pos}
\newcommand{\PoS}{\operatorname{PoS}}

Spiegelhalter and Freedman also proposed the use of the \textit{``probability of concluding that the new treatment is superior and of this being correct''} ($P_{Ss}$ in their notation) to derive a required sample size~\citep{spiegelhalter1986}.
The quantity has also been referred to as `prior adjusted power'~\citep{spiegelhalter-2004,shao-2008}.
This definition of probability of success is also discussed in \citet{liu-2010} and \citet{ciarleglio-2015}.
In the situation at hand, it reduces to
\begin{align}
    \PoS(n) :=&\ \Pr_{}[\,Z_n > \criticalvalue, \Theta \geq \mcid\,] \\
          =&\ \int_{\mcid}^\infty \int_{\criticalvalue}^\infty \phi(z - \theta_n)\ \varphi(\theta)\ \operatorname{d}z\operatorname{d}\theta  \ ,
\end{align}
where $\phi$ is the PDF of the standard normal distribution.
Here, we are slightly more general than previous authors in that
we allow $\mcid>0$ and use a tighter definition of `success': a trial is only successful if the null hypothesis is rejected \emph{and} the effect is relevant.
Whenever $\mcid=0$ this coincides with the definitions used previously in the literature.

The definition of $\PoS(n)$ critically relies on what is being considered a `success'.
The original proposal of Spiegelhalter and Freedman only considers a significant study result a success if the underlying effect is also non-null (i.e., the joint probability of non-null \emph{and} detection).
In more recent publications, a majority of authors tend to follow
O'Hagan~\textit{et~al.} who consider a slightly
different definition of the probability of success by integrating the probability to reject over the entire parameter range~\citep{o2001bayesian,ohagan-2005}
and term this `assurance'.
For a more comprehensive overview of the terms used in
the literature, see Section~\ref{sec:literature}.
The alternative definition for probability of success introduced by O'Hagan~\textit{et~al.}
corresponds to the marginal probability of rejecting the null hypothesis irrespective of the corresponding parameter value
\begin{align}
    \PoS'(n) :=&\ \Pr_{}[\,Z_n > \criticalvalue\,]  \\
           =&\ \int_{-\infty}^\infty \int_{\criticalvalue}^\infty \phi(z - \theta_n)\ \varphi(\theta)\ \operatorname{d}z\operatorname{d}\theta\\
           =&\ \PoS(n) + \underbrace{\Pr_{}[\,Z_n > \criticalvalue, 0<\Theta<\mcid\,]}_{\text{probability of rejection and irrelevant effect}} + \underbrace{\Pr_{}[\,Z_n > \criticalvalue, \Theta \leq 0\,]}_{\text{probability of a type~I error}} \ .
           \label{eq:pos-prime-decomposition}
\end{align}
The decomposition in equation~\eqref{eq:pos-prime-decomposition} makes it clear that the implicit definition of `success' underlying $\PoS'(n)$ is at least questionable \citep{liu-2010}.
The marginal probability of rejecting the null hypothesis includes rejections under irrelevant or even null values of $\theta$, and is thus inflated by type~I errors and rejections under irrelevant values of~$\theta$.
This issue was first raised by \citet{spiegelhalter-2004}
for simple (point) null and alternative hypotheses.
The degree to which $\PoS(n)$ and $\PoS'(n)$ differ numerically is discussed in more detail in Section~\ref{sec:results:pos-for-sample-size-determination}.
Which definition of `success' is preferred mainly depends on perspective: a short-term oriented pharmaceutical company might just be interested in rejecting the null hypothesis to monetise a new drug - irrespective of it actually showing a relevant effect. This view would then correspond to $\PoS'(n)$.
Regulators and companies worried about the longer-term consequences of potentially having to retract ineffective drugs,
might tend towards the joint probability of correctly rejecting the null, i.e.,~$\PoS(n)$.
We take the latter perspective and focus on~$\PoS(n)$.

$\PoS(n)$ is an unconditional quantity and must therefore implicitly depend on the \textit{a~priori} probability of a relevant effect.
To see this, consider the following decomposition
\begin{align}
    \PoS(n) &=
        \Pr_{}[\,Z_n > \criticalvalue, \Theta \geq \mcid\,] \\
        &= \int_{\mcid}^\infty \Pr_\theta[\,Z_n > \criticalvalue\,]\ \varphi(\theta)\ \operatorname{d} \theta \label{eq:pos-integral} \\
        &= \Pr_{ }[\,Z_n > \criticalvalue \mid \Theta \geq \mcid\,] \, \Pr_{}[\,\Theta \geq \mcid\,] \\
        &= \underbrace{\E\big[\, \Pr_{\, \Theta \geq \mcid}[\,Z_n > \criticalvalue\,] \,\big]}_{= \ \E[\,\rpower{n}\,]\ =:\ \EP(n)} \,
        \Pr_{}[\,\Theta \geq \mcid\,]  \ .
        \label{eq:pos-and-ep}
\end{align}
This means that the probability of success can be expressed as the product of the `expected power', $\EP(n)$, and the \textit{a~priori} probability of a relevant effect (see again \citet{spiegelhalter-2004} for the situation with point hypotheses).
Expected power was implicitly mentioned in \citet{spiegelhalter1986} ($P_{Ss}/P_{\cdot s}$ in their notation) as a way to characterise the properties of a design.
The use of expected power as a means to derive the required sample size of a design under uncertainty was then proposed in \citet{brown1987projection} by solving
\begin{align}
    {n_{\scalebox{.5}{EP}}^*} :=\quad &\argmin{n} &&n \nonumber\\
                      &\st &&\EP(n) \geq 1 - \beta \ . \label{eq:ep-problem}
\end{align}
Since the power function is monotonically increasing in $\theta$, expected power is strictly larger than power at the minimal relevant value whenever $\Pr[\,\Theta>\mcid\,] > 0 $.
This implies that a constraint on expected power instead of a constraint on the probability to reject the null hypothesis at $\mcid$ is less restrictive, and consequently~${n_{\scalebox{.5}{EP}}^*}<{n^*_{\mcid}}$.

The terms `expected power' and `probability of success' are sometimes used interchangeably in the literature (see Section~\ref{sec:literature}).
In the following, we take a closer look at their connection to clarify their characteristic differences.
Expected power is merely a weighted average of the probability to reject in the relevance region $\theta\geq\mcid$, where the weight function is given by the conditional prior density  $\varphi(\cdot\,|\,\Theta\geq \mcid)$ defined in equation~\eqref{eq:conditional-prior-density}
\begin{align}
    \EP(n) =&\ \E\big[\, \Pr_{\Theta}[\,Z_n >\criticalvalue\,] \mid \Theta > \mcid\,\big] \\
        =& \int_{\mcid}^\infty \Pr_\theta[\,Z_n > \criticalvalue\,]\ \varphi(\theta\,|\,\Theta\geq \mcid)\ \operatorname{d} \theta \ .
        \label{eq:ep-integral}
\end{align}
$\PoS(n)$, on the other hand, integrates the probability to reject over the same region using the unconditional prior density (see equations~\eqref{eq:pos-integral} and \eqref{eq:ep-integral}).
Thus, in~contrast to $\PoS(n)$, expected power does not depend on the \textit{a~priori} probability of a relevant effect size but only on the relative magnitude of the prior density (`\textit{a~priori} likelihood') of relevant parameter values.
Since the conditional prior density differs from the unconditional one only by normalisation via the \textit{a~priori} probability of a relevant effect,
it follows from equation~\eqref{eq:pos-and-ep} that $\EP(n)$ and $\PoS(n)$ differ only by the constant factor $ \Pr_{}[\,\Theta \geq \mcid\,]$.
Consequentially, any constraint on probability of success can be reformulated as a constraint on expected power and \textit{vice~versa}
\begin{align}
    \PoS(n)\geq 1 - \beta
    \quad \Leftrightarrow \quad
    \EP(n)\geq (1 - \beta)\,/\,\Pr_{ }[\,\Theta \geq \mcid\,] \ .
    \label{eq:pos-to-ep-cnstr}
\end{align}
Furthermore, since $\PoS(n) = \EP(n)\Pr_{}[\,\Theta \geq \mcid\,]$ and $\EP(n)\leq 1$, $\PoS(n)$ can never exceed the \textit{a~priori} probability of a relevant effect, $\Pr_{}[\,\Theta \geq \mcid\,]$.
This implies that the usual conventions on the choice of $\beta$
as the maximal type~II error rate for a point alternative
cannot be meaningful in terms of the unconditional $\PoS(n)$, since the maximum attainable probability of success is the situation-specific \textit{a~priori} probability of a relevant effect.
The need to recalibrate typical benchmark thresholds when considering probability of success was previously discussed in the literature.
For instance, \citet{o2001bayesian} states that
``[t]he assurance figure is often much lower [than the power], because there is an appreciable prior probability that the treatment difference is less than $\delta^*$'', where in their notation, $\delta^*$ corresponds to $\mcid$ in our notation.
A similar argument is put forward in \citet[Section~2]{rufibach_15} for $\PoS'(n)$.
The key issue is thus whether one is interested in the joint probability of rejecting the null hypothesis \emph{and}
the effect being relevant,~$\PoS(n)$, or
the conditional probability of the rejecting the null hypothesis \emph{given} a relevant effect,~$\EP(n)$.
While the interpretation of both quantities is different, in any particular situation, they only differ by a constant factor.

\subsection{Expected power versus quantile-based approach for sample size derivation}

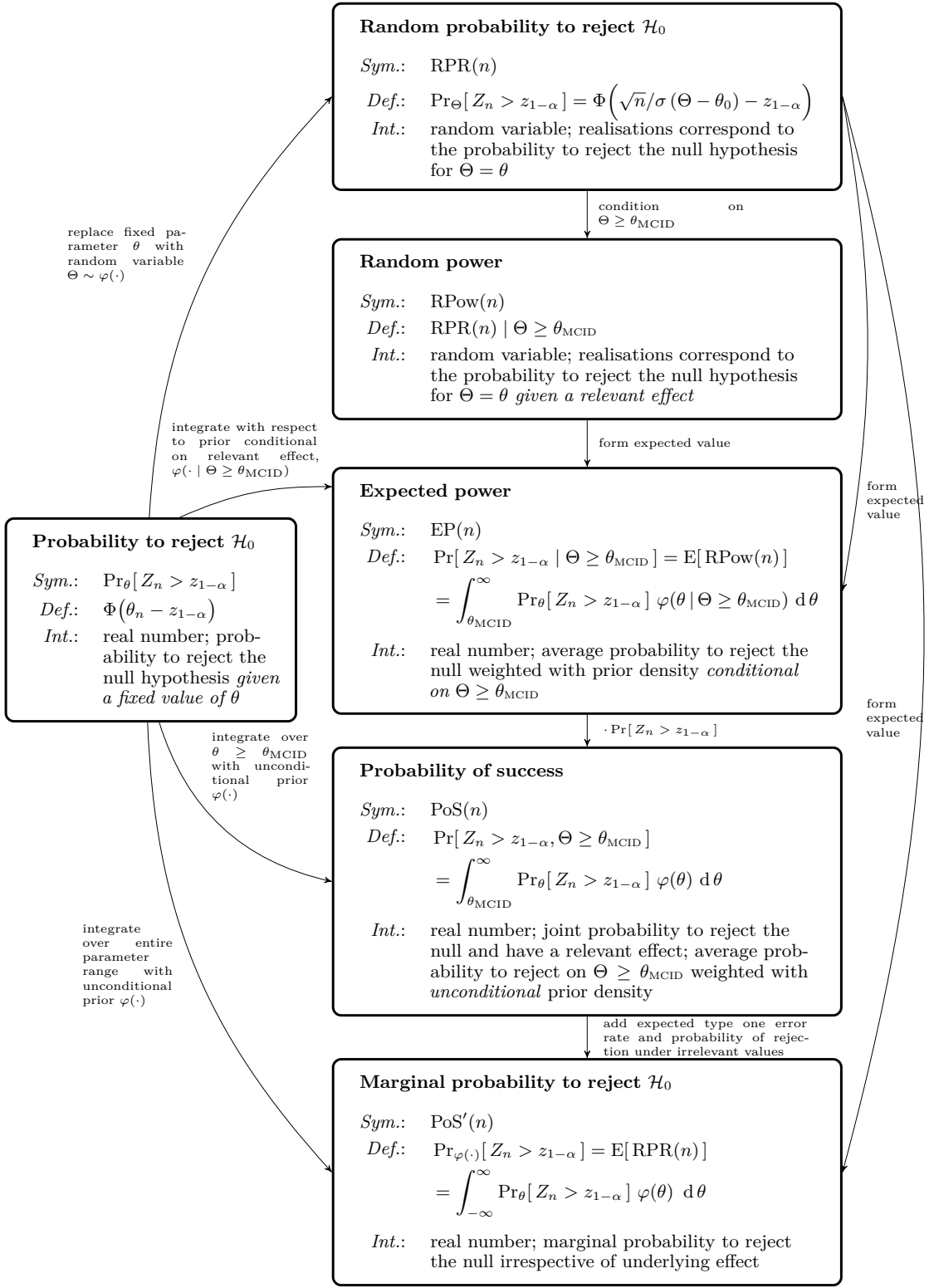
\begin{figure}
\centering
\resizebox{0.925\textwidth}{!}{\input{diagram}}
\caption{Structured overview of all quantities related to `power' that are introduced in Sections~\ref{sec:intro} to~\ref{sec:pos}. The symbols used in the text (\textit{Sym.}), their exact definitions (\textit{Def.}), and verbal interpretation (\textit{Int.}) are summarised in the respective boxes. The relationships between the individual quantities are given as labelled arrows.
For an overview of previous mentions and synonyms used in the literature, see Table~\ref{tab:literature-overview}.}
\label{fig:diagram}
\end{figure}

Since expected power and probability of success are proportional,
it suffices to compare expected power and the quantile-based approach outlined in Section~\ref{sec:quantile-approach} with respect to sample size derivation.
Consider
${\theta'' := \quant{1 - \gamma}[\, \Theta \geq \mcid \,]}$ and an arbitrary but fixed parameter value $\theta'>\theta'' >\mcid$.
Clearly, under the quantile-based approach, the rejection probability  at any $\theta'> \theta''$ does not contribute towards the fulfilment of the power constraint since the probability to reject is only evaluated at
${\theta''}$.
For expected power, however, the total functional derivative with respect to changes in the probability to reject at $\theta'$ and $\theta''$ is
\begin{align}
    \operatorname{d} \operatorname{EP}(n) &=
        \frac{\partial\operatorname{EP}(n)}{\partial\Pr_{\theta''}[\,Z_n > \criticalvalue\,]}\operatorname{d}\Pr_{\theta''}[\,Z_n > \criticalvalue\,] +
        \frac{\partial\operatorname{EP}(n)}{\partial\Pr_{\theta'}[\,Z_n > \criticalvalue\,]}\operatorname{d} \Pr_{\theta'}[\,Z_n > \criticalvalue\,] \\[1em]
        &= \varphi(\theta''\mid\theta \geq \mcid)\operatorname{d} \Pr_{\theta''}[\,Z_n > \criticalvalue\,] +
        \varphi(\theta'\mid\theta \geq \mcid)\operatorname{d} \Pr_{\theta'}[\,Z_n > \criticalvalue\,] \ .
\end{align}
Keeping expected power constant, i.e., setting ${\operatorname{d} \operatorname{EP}(n)=0}$ and solving for $\operatorname{d}\Pr_{\theta'}[\,Z_n > \criticalvalue\,]$  yields
\begin{align}
    0 &= \varphi(\theta''\mid\Theta \geq \mcid)\operatorname{d} \Pr_{\theta''}[\,Z_n > \criticalvalue\,] +
        \varphi(\theta'\mid\Theta \geq \mcid)\operatorname{d} \Pr_{\theta'}[\,Z_n > \criticalvalue\,] \\[1em]
    \Leftrightarrow \operatorname{d} \Pr_{\theta'}[\,Z_n > \criticalvalue\,] &= -\frac{\varphi(\theta'\mid\Theta \geq \mcid)}{\varphi(\theta''\mid\Theta \geq \mcid)} \operatorname{d} \Pr_{\theta''}[\,Z_n > \criticalvalue\,] \\[1em]
    \Leftrightarrow \operatorname{d} \Pr_{\theta'}[\,Z_n > \criticalvalue\,] &= -\frac{\varphi(\theta')}{\varphi(\theta'')} \operatorname{d} \Pr_{\theta''}[\,Z_n > \criticalvalue\,]
    \ .
\end{align}
A reduction in the probability to reject at $\theta''$ by one percentage point can thus be compensated by an increase in the probability to reject at $\theta'$ by~${\varphi(\theta')\,/\,\varphi(\theta'')}$ percentage points.
This demonstrates that the core difference between the prior quantile-based approach and the expected power approach is whether or not a trade-off between power at different parameter values is deemed permissible (expected power) or not (quantile-based approach).
A structured overview of the terms introduced so far and the respective connections between them is given in Figure~\ref{fig:diagram}.

\section{Connection to utility maximisation}
\label{sec:utility}

In a regulatory environment, and most scientific fields,
the choice of the significance level, $\alpha$, is a pre-determined
quality criterion.
In the life sciences a one-sided $\alpha$ of 2.5\% is common.
Yet, the exact choice of the threshold $1-\beta$ is much more arbitrary.
In clinical trials, $1-\beta=0.9$ or $1-\beta=0.8$ are common choices when
a classical sample size derivation is conducted.
From the previous section it is already clear that a
generic threshold for $1-\beta$ that is independent of the specific context of a trial only makes sense with conditional approaches like the (conditional prior) quantile approach or when using $\EP(n)$ to derive a required sample size.
In principle, the unconditional $\PoS(n)$ should be easier to interpret
by non-statisticians.
Equation~\eqref{eq:pos-to-ep-cnstr} allows the transformation of an $\EP(n)$-based
sample size derivation, which can readily use any of the established values for $1-\beta$, into a $\PoS(n)$-based sample size derivation by
re-calibrating the threshold with the proportionality factor linking $\EP(n)$ and $\PoS(n)$.
This only transforms the conditional criteria (minimum $\EP(n)$)
of the classical sample size derivation to the unconditional domain (minimum $\PoS(n)$) without affecting the derived sample size in any way.
For instance, if $1-\beta=0.8$ and $\Pr[\,\Theta\geq\mcid\,] = 0.683$,
the transformed threshold for $\PoS(n)$ would be~$0.5464$.
In making the assumptions underlying the sample size derivation more transparent by formulating them in terms of unconditional probabilities, there is a need to explain to practitioners why the threshold
now differs from study to study.

The Bayesian view and the prior density $\varphi(\cdot)$ give a natural
answer to this issue of trial-specific thresholds via
the concept of utility maximisation or maximal expected utility (MEU).
An in-depth discussion of the MEU concept is beyond the scope of
this manuscript and we refer the reader to, for example, ~\citet{lindley-1997}.
We merely want to highlight the fact that the choice of the constraint threshold $1-\beta$ can be justified by making the link to MEU principles.
To this end we consider a particularly simple utility function.

Assume that the maximal type~I error rate is still to be controlled at level $\alpha$.
For sake of simplicity, further assume that a \emph{correct}
rejection of the null hypothesis yields an expected return of $\lambda$.
Here the return is given in terms of the average per-patient costs within the trial.
Ignoring fixed costs, the expected trial utility (in units of average per-patient costs) is then given by
\begin{align}
    U(n) := \lambda\,\PoS(n) - n
    \label{eq:utility-function}
\end{align}
and the utility-maximising sample size is $n_{U}^*(\lambda) := \argmax{n} U(n)$.
Obviously, the same sample size would be obtained by solving problem \eqref{eq:ep-problem}, given
\begin{align}
    1-\beta = \EP\big( n_{U}^*(\lambda) \big) = \Pr_{ }[\, \Theta \geq \mcid \,] \, \PoS\big(n_{U}^*(\lambda)\big) \ .
\end{align}
The right hand side, $\Pr_{}[\, \Theta \geq \mcid \,] \, \PoS\big(n_{U}^*(\lambda)\big)$, is the utility-maximising expected power threshold given the utility parameter~$\lambda$.
Similarly, one could start with $n_{\EP}^*$ for a given $\beta$ and derive the corresponding $\lambda$ such that $n_{U}^*(\lambda)=n_{\EP}^*$.
This value of $\lambda$ would then correspond to the implied expected reward upon successful rejection of the null for given $\beta$.
Under the assumption of a utility function of the form (\ref{eq:utility-function}), $\lambda$ and $\beta$ can thus be matched such that the corresponding utility maximisation problem and the constraint minimisation of the sample size under a power constraint both lead to the same required sample size.
Consequently, practitioners are free to either define an expected return upon successful rejection, $\lambda$, or a threshold on the minimal expected power, $1-\beta$.

While it is theoretically attractive to derive the sample size directly based on a utility function, an informed choice of $\lambda$ is often hard to justify in practice.
In these situations one may instead reverse the perspective and determine the value of $\lambda$ under which the utility-maximising design would coincide with the design obtained under a standard (expected) power threshold of, say, 80\% or 90\%.
The implied reward parameter might then be used to communicate the consequences of different choices of power thresholds to decision makers and to inform the final choice of $1-\beta$.
This approach can, of course, be generalised to more detailed utility functions.
Note, however, that for utility functions with more than one free parameter there is no longer a one-to-one correspondence between power level and utility parameters.
Rather, for a given power level, there will be a level-set of values for the utility parameters that match the specified power.
We give a practical example of this process in Section~\ref{sec:example:clinical-trial}.

\begin{center}
\renewcommand{\arraystretch}{1.5}
\begin{longtable}{p{3cm}p{4cm}p{8cm}}
    \caption{%
        Selected publications on `hybrid' sample size derivation based on error rates. Structured by concepts as defined in Figure~\protect\ref{fig:diagram}.
    }%
    \label{tab:literature-overview}\\

    \textbf{Concept} & \textbf{References} & \textbf{Notes} \\
    \hline
    \endfirsthead

    \input{literature-review-table}

\end{longtable}
\end{center}

\section{Literature review of terminology}
\label{sec:literature}

A structured overview of the literature on `hybrid' Bayesian sample size derivation in the context of clinical trials is given
in Table~\ref{tab:literature-overview}.
The table relates publications in the field to the terms defined
in Figure~\ref{fig:diagram}.
Publications with a similar take on the matter are grouped.
In the following, we highlight a few particularly interesting contributions and how they relate to the definitions used in this manuscript.

The majority of the manuscripts only consider
the marginal probability to reject $\mathcal{H}_0$ ($\PoS'(n)$).
Many publications refer to \citet{o2001bayesian} or \citet{ohagan-2005},
where this quantity was introduced as `assurance'.
The range of names for what we call the `marginal probability to reject $\mathcal{H}_0$' is, however, quite diverse:
`assurance', `probability of success',
`predictive probability of success', `average probability of success', `probability of statistical success', `probability of study success', `predictive power', `predictive frequentist power', `expected power', `average power', `strength', `extended Bayesian expected power 1', and `hybrid Neyman-Pearson-Bayesian probability'.

However, only a handful of authors elaborate on the intricacies of defining what exactly constitutes a `success' and
whether to consider an unconditional measure of success or to condition on the presence of a relevant effect for sample size derivation \citep{spiegelhalter1986,brown1987projection,shao-2008,liu-2010,ciarleglio-2015}.
Most publications fail to define explicitly what exactly constitutes a `success'.
Yet, the use of $\PoS'(n)$ implies that rejection of the null hypothesis,
irrespective of its truth, must be considered a success.
Our analysis in Section~\ref{sec:results:pos-for-sample-size-determination}
confirms the statement in \citet{spiegelhalter-2004} that
$\PoS'(n)$ can be used as a practical approximation to
$\PoS(n)$ in many situations.
The exact definition of `probability of success' becomes more
interesting when allowing for $\mcid>\theta_0$,
a potential extension rarely considered in the literature
(see, e.g.,~\citet{brown1987projection} for the binary case).
We revisit the distinction between $\PoS'(n)$ and $\PoS(n)$
in a concrete example in Section~\ref{sec:results:pos-for-sample-size-determination}.

The exact choice of wording should not be given too much weight.
However, we feel that any notion of power in the `hybrid' Bayesian/frequentist setting should be \emph{conditional} on a relevant effect (or at least a non-null effect) to preserve the conditional nature of the purely frequentist power.
Using the term `power' to refer to a joint probability like
the `expected power' of
\citet{brown1987projection} and \citet{ciarleglio-2015} (our $\PoS(n)$) or
the `average/expected power' of
\citet{spiegelhalter-2004} (our $\PoS'(n)$) is potentially misleading.
Others suggest `conditional expected power' for $\EP(n)$ to distinguish it from `expected power' (our $\PoS'(n)$)~\citep{brown1987projection,ciarleglio-2015}.
This wording, however, may lead to confusion when also considering interim analyses where `conditional power' is a well-established term for the probability of rejecting the null hypothesis given $\altv$ and partially observed data~\citep{bauer-2016}.

A particularly interesting publication is
\citet{liu-2010}.
They extend hybrid sample size derivation in the normal case to also incorporate uncertainty about the
variance and clearly distinguish between
$\PoS'(n)$ = `extended Bayesian expected power 1',
$\PoS(n)$ = `extended Bayesian expected power 2',
and $\EP(n)$ = `extended Bayesian expected power 3'.
Apart from nomenclature, our definitions of these three quantities
only differ in that they assume the standard deviation to be
fixed and the fact that we accommodate the optional notion of a relevant effect via $\mcid$.
The former makes explicit formulas more manageable,
the latter is important to keep sample sizes small in situations with vague or conservative prior information but substantial relevance thresholds.
\citet{liu-2010} and \citet{rufibach_15} are also the only
publications we found that study the distribution of the quantities that are averaged over ($\rpr(n)$ and $\rpower{n}$ in our notation, see Figures~\ref{fig:power-distribution} and~\ref{fig:power-distribution-quantile-approach}).
In \citet{ciarleglio-2015}, the distinction between all three
quantities is also made explicit (`expected power' is our $\PoS'(n)$, `prior-adjusted power' is our $\PoS(n)$, and `conditional expected power' is our $\EP(n)$).

\section{Prior elicitation}

A major issue in modelling uncertainty and computing sample size via prior densities is the elicitation of an adequate prior.
At first glance, non-informative or `objective' priors seem to be a viable choice.
As illustrated in \citet{rufibach_15}, the prior crucially impacts the properties and interpretability of $\PoS'$ and likewise any other quantity depending on $\varphi(\cdot)$ in Figure~\ref{fig:diagram}, so careful selection is paramount.
Often in clinical research, there is no direct prior knowledge on the effect size of interest, e.g., overall survival in a phase III trial, as no randomised trials comparing these same treatments using the same endpoint have been run previously.
Researchers are then often tempted to use a {\it vague} prior, typically a normal prior with large variance, as, e.g., advocated in \cite{saint-hilary-2019}.

Assuming a non-informative, improper prior for $\Theta$ would imply that arbitrarily large effect sizes are just as likely as small ones.
Yet, in clinical trials, the standardised effect size rarely exceeds 0.5 \citep{lamberink2018}.
We thus illustrate the characteristics of the different approaches to defining power constraints under uncertainty using a convenient truncated Gaussian prior.
The truncated Gaussian is conjugate to a Gaussian likelihood and allows us to restrict the plausible range of effect sizes to, e.g., liberally $[-1, 1]$.
Also, the truncated Gaussian is the maximum entropy distribution on the truncation interval, given mean and variance which can be interpreted as a `least-informative' property under constraints on the first two moments.

Alternatively, a trial designer can formally elicit a prior on the effect size of interest.
Kinnersley and Day describe a structured approach to the elicitation of expert beliefs in a clinical trial context based on the SHELF framework~\citep{kinnersley_13,shelf}.
\citet{dallow_18} discusses how SHELF is routinely used within a pharmaceutical company to define prior distributions that are used in conjunction with calculation of probability of success and to inform internal decision making at key project milestones.
Formal and informal prior elicitation is also discussed in~\citet{spiegelhalter-2004}.

\section{Results}

\subsection{Comparison of required sample sizes for various prior choices}

\begin{figure}[htbp]
    \centering
    \includegraphics[width=\textwidth]{%
        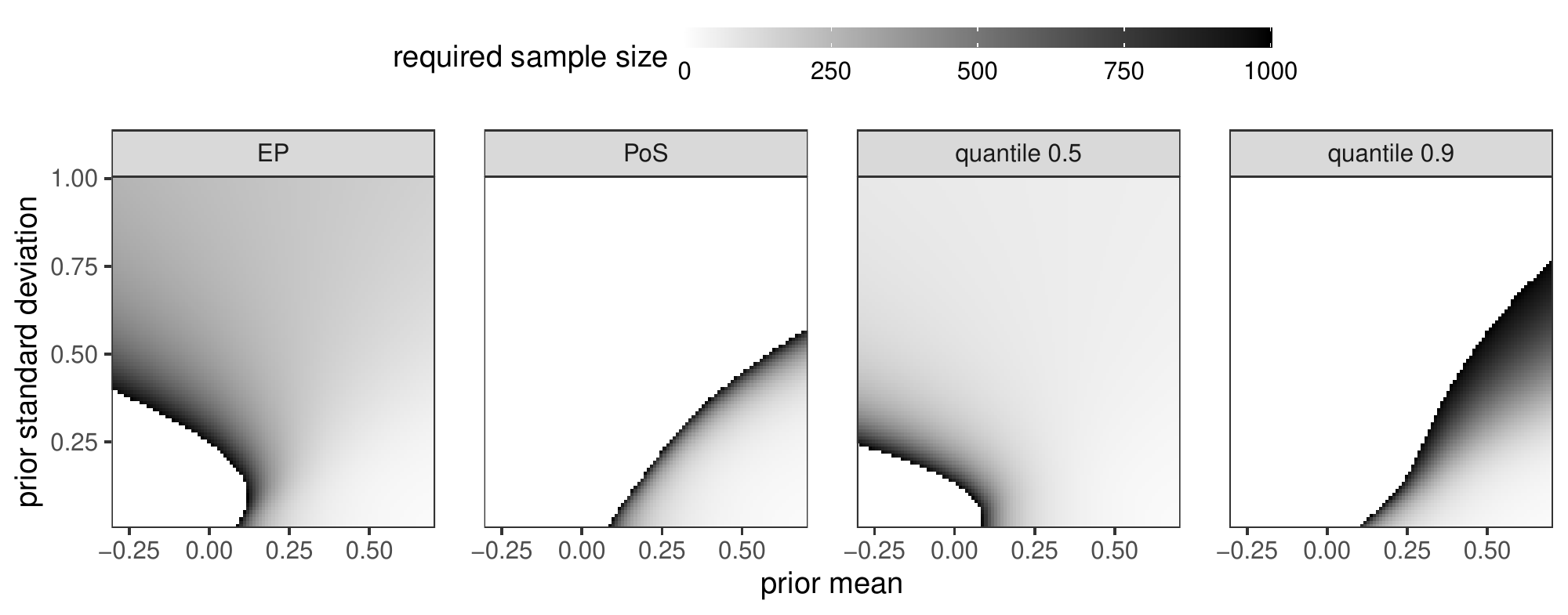}
    \caption{Required sample size plotted against prior parameters (Normal truncated to [-0.3, 0.7], with varying mean and standard deviation); $\mcid=0.1$; EP = Expected Power, PoS = Probability of Success, quantile = quantile approach with $\gamma=0.5$ and $\gamma=0.9$, respectively.}
    \label{fig:n-vs-prior-parameters}
\end{figure}

Let $\mcid=0.1$ and let the maximal feasible sample size be 1000.
Figure~\ref{fig:n-vs-prior-parameters} shows the required sample sizes under the expected power, the probability of success, and the quantile approach (for $\gamma = 0.5, 0.9$).
We use $\alpha = 0.025$ and $\beta = 0.2$.

The patterns of required sample sizes are qualitatively different between the three approaches.
For probability of success, large prior uncertainty implies low \textit{a~priori} probability of a relevant effect and thus the required sample sizes explode for large prior standard deviations (in relation to the prior mean).
For very large standard deviations, the constraint on probability of success becomes infeasible.

The expected power criterion leads to a completely different sample size pattern.
Since expected power is defined conditional on a relevant effect, large prior uncertainty increases the weight in the upper tails of the power curve where power quickly approaches one.
Consequently, for small prior means, larger uncertainty decreases the required sample size.
For large prior means, however, smaller prior uncertainty leads to smaller sample sizes since again more weight is concentrated in the tails of the power curve.

The characteristics of the prior-quantile approach
very much depend on the choice of $\gamma$.
When using the conditional prior median ($\gamma=0.5$) the approach is qualitatively similar to the expected power approach.
This is due to the fact that computing power on the conditional median of the prior is close to computing power on the conditional prior mean.
Since the power function is locally linear around the centre of mass of the conditional prior, this approximates computing expected power by interchanging forming the expected value and computing power (i.e., first average the prior and then compute power or average over power with weights given by the conditional prior).
For a stricter criterion ($\gamma=0.9$) the required
sample sizes are much larger.
This is due to the fact that the quantile approach
does not allow a trade-off between power in the upper tails of the power curve and regions with low power. Higher uncertainty then decreases the $(1-\gamma)$-quantile towards the minimal relevant effect and thus increases the required sample size.

\subsection{Probability of success as the basis for sample size derivation}
\label{sec:results:pos-for-sample-size-determination}

In theory, one might be inclined to derive a sample size based on the probability of success instead of using expected power.
Consider a situation in which the \textit{a~priori} probability of $\Theta \geq \mcid$ is $0.51$.
The probability of success is then only $41\%$ (for 80\% expected power) or $46\%$ (for 90\% expected power).
A sponsor might want to increase these relatively low unconditional success probabilities by deriving a sample size based on a minimal $\PoS(n)$ of $1-\beta$ instead.
The choice of $1-\beta$ is limited by the \textit{a~priori}
probability of a relevant effect (0.51 in this case).
Using equation~\eqref{eq:pos-to-ep-cnstr} a minimal probability of success of 0.5 is equivalent to requiring an expected power of more than 98\%.
In essence, the attempt to increase $\PoS(n)$ via a more stringent threshold on $\EP(n)$ implies that low \textit{a~priori} chances of success are to be offset with an almost certain detection ($\EP(n)\approx 1$) in the unlikely event of an effect actually being present.
The ethical implication of this approach to sample size derivation is that an extremely large number of patients would be exposed to treatment although the sponsor more or less expects it to be ineffective.

This example demonstrates that the (situation agnostic) conventions on power levels cannot be transferred to thresholds for probability of success without adjustment for the situation-specific \text{a~priori} probability of a relevant effect.
It thus seems much easier to directly impose a constraint on expected power, which implicitly adjusts for the \textit{a~prior} probability of a relevant effect via equation~\eqref{eq:pos-to-ep-cnstr}.

\begin{figure}
    \centering
    \includegraphics[width=\textwidth]{%
        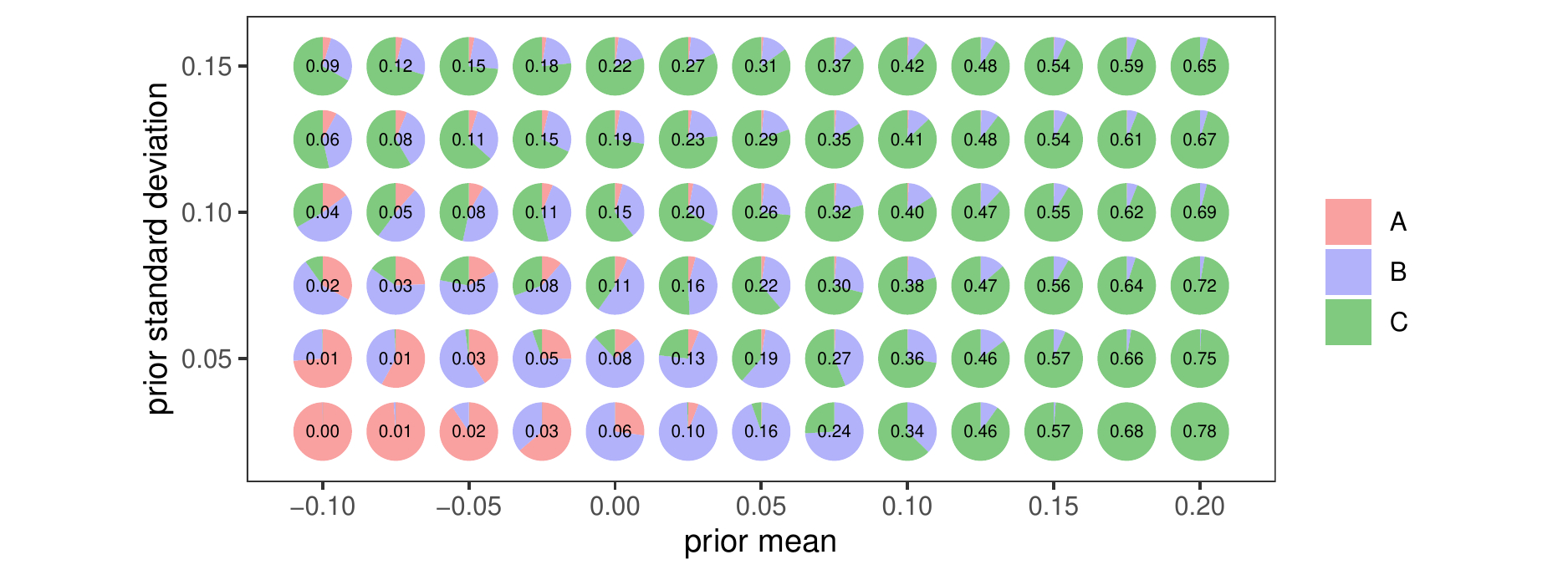}
    \caption{Components of $\PoS'(n)$ for $n=150$, $\theta_0=0$, $\alpha=0.25$, $\mcid=0.1$ and varying prior mean and standard deviation;
    numbers correspond to overall $\PoS'(n)$;
    proportions in individual pie charts correspond to:
    A = probability to reject and null effect (type~I error),
    B = probability to reject and irrelevant but non-null effect,
    C = probability to reject and relevant effect ($\PoS$).}
    \label{fig:pos-components}
\end{figure}

To investigate the difference between the probability of success, $\PoS(n)$, and the marginal probability of rejecting the null hypothesis, $\PoS'(n)$, Figure~\ref{fig:pos-components} visualises the proportion of the individual components of $\PoS'(n)$ for varying prior standard deviation and prior means.
The sample size is fixed at $n=150$, $\theta_0=0$, the maximal type~I error rate is $\alpha=0.025$, and the minimal clinically important difference is $\mcid=0.1$.
Evidently, the contribution of type~I errors (component `A' in Figure~\ref{fig:pos-components}) to $\PoS'(n)$ is mostly negligible unless the prior is sharply peaked at an effect size slightly smaller than the null.
The \textit{a~priori} probability of a relevant effect size is close to zero in these cases and so is $\PoS'(n)$.
For the more practically relevant scenarios with prior mean greater than $\theta_0=0$, the contribution of the average type~I error rate to $\PoS'(n)$ is almost negligible.
Still, if $\mcid>\theta_0$, $\PoS'(n)$ might be inflated substantially by rejections under parameter values that are non-null but also not clinically relevant.
This phenomenon evidently depends on the magnitude of $\mcid$; the more of the prior mass concentrated in $[\,0,\mcid\,]$, the larger the contribution towards $\PoS'(n)$.
If $\mcid = 0$, the numeric difference between $\PoS$ and $\PoS'$ is negligible since the maximal type~I error rate is controlled at level $\alpha$ and the power curve quickly approaches zero on the interior of the null hypothesis.
This was already pointed out by \citet{spiegelhalter-2004},
who argue that $\PoS'(n)$ can be used as approximation to $\PoS(n)$ in many practically relevant situations.

\subsection{Distribution of random power under a constraint on expected power}

\begin{figure}
    \centering
    \includegraphics[width=\textwidth]{%
        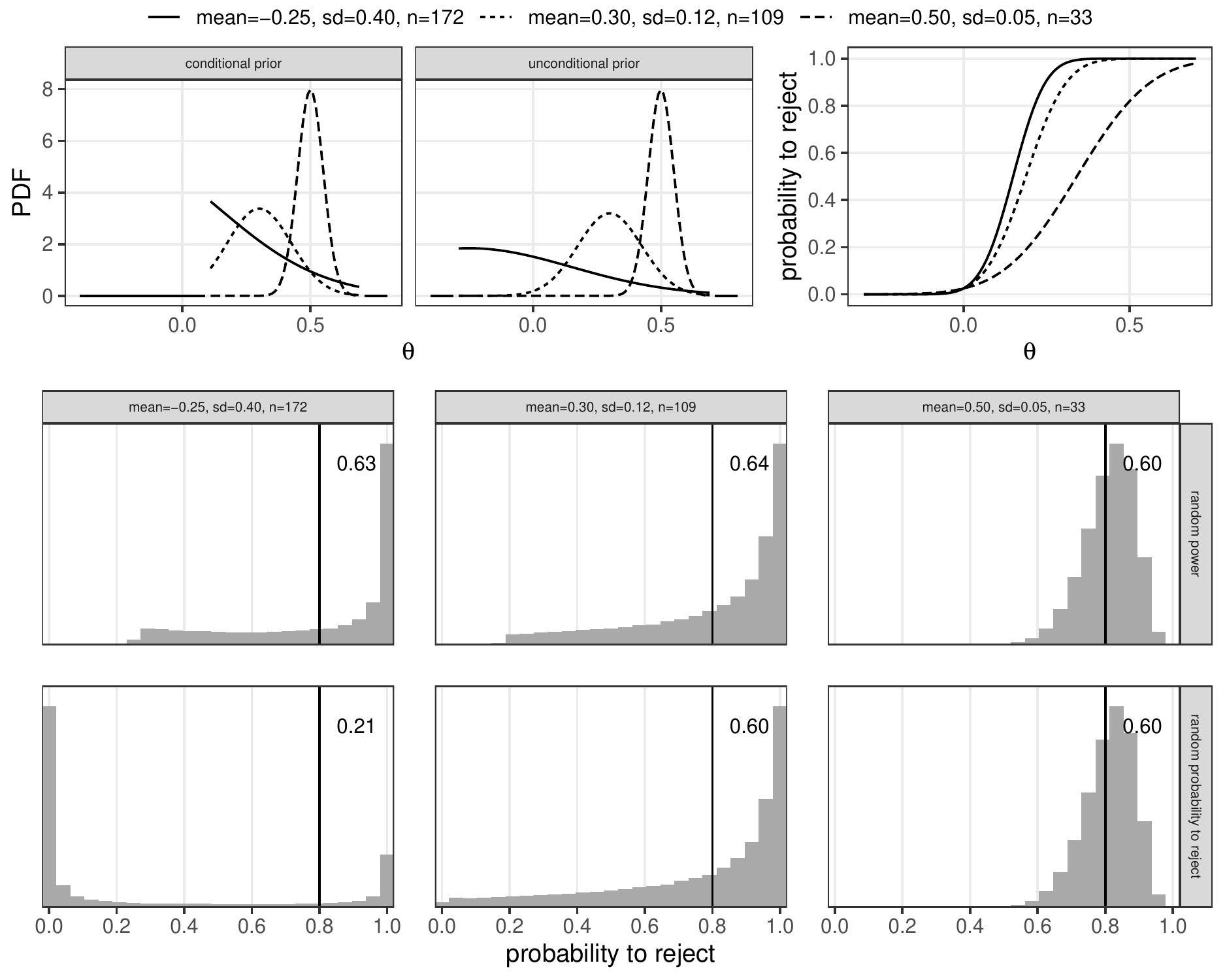}
    \caption{%
        Top: conditional and unconditional prior PDF and power curves corresponding to $n_{\protect\scalebox{.5}{EP}}^*$ for $1-\beta=0.8$.
        Bottom: histogram of the random power (conditional on relevant effect) and the random probability to reject (unconditional); vertical lines indicate 80\% power and the numbers are the respective (conditional) probabilities to exceed a probability to reject of 80\%.
    }
    \label{fig:power-distribution}
\end{figure}

To further investigate the properties of the random variable $\Pr_{\Theta}[\,Z_n>\criticalvalue\,]$, we consider three example prior configurations with means $-0.25, 0.3, 0.5$ and standard deviations $0.4, 0.125, 0.05$ respectively.
The corresponding sample sizes to reach an expected power of at least 80\% are 854, 126, and 32.
Figure~\ref{fig:power-distribution} shows the unconditional and conditional (on a relevant effect) priors, the corresponding probability of rejecting the null as a function of $\theta$, and histograms of the distributions of random power ($\rpower{n}$), and the unconditional probability to reject the null hypothesis ($\rpr(n)$).

Clearly, the distributions of the conditional and unconditional rejection probabilities (random power and random probability to reject, respectively) are qualitatively very different in the three situations.
In the first case (mean $-0.25$, standard deviation 0.4), the prior mass is mostly concentrated on the null hypothesis and the normalising factor that links the unconditional and the conditional prior is clearly noticeable.
The conditional prior then assigns most weight to values of $\theta$ close to the relevance threshold leading to
a large required sample size.
The large sample size then implies a steep power curve and a distribution of $\rpower{n}$ that is highly right-skewed towards~1 since $\rpower{n}$ is conditional on $\Theta>\mcid$.
If the unconditional distribution of the rejection probability is considered instead ($\rpr(n)$), the characteristic u-shape discussed in \citet{rufibach_15} is recovered.

For the intermediate setting (mean 0.3, standard deviation 0.125), most prior mass is already concentrated on relevant values.
The difference between conditional and unconditional prior is less pronounced (the normalising factor is closer to~1) and even the unconditional distribution of the rejection probability is no longer u-shaped.

Finally, in the last setting (mean 0.5, standard deviation 0.05), $\theta$ is almost certainly highly relevant.
Since the normalising factor is thus close to~1, there is no discernible difference between the conditional and the unconditional prior densities.
Not surprisingly, the assumption of a highly relevant effect with high certainty only requires a small sample size (32).
This leads to a relatively flat curve of the rejection probability and to a peaked distribution of $\rpower{n}$ and $\rpr(n)$.
Flat power curves and high certainty about the effect size tend to result in peaked distributions of $\rpower{n}$ and $\rpr(n)$ because the power curve is almost linear in the region of the parameter where the prior mass is concentrated.
The distribution of $\rpower{n}$ and $\rpr(n)$ is thus well approximated by a linear transformation of the (conditional) prior, which is a peaked truncated normal distribution.
Since conditioning has almost no effect, the unconditional distribution of the probability to reject is the same in this case.

Interestingly, both settings with higher \textit{a~priori} uncertainty lead to a high chance of exceeding a power of 80\%.
This is due to the fact that the rare occurrence of very low rejection probabilities needs to be compensated to achieve an overall expected power of 80\%.

\subsection{Distribution of random power under quantile approach}

\begin{figure}
    \centering
    \includegraphics[width=.99\textwidth]{%
        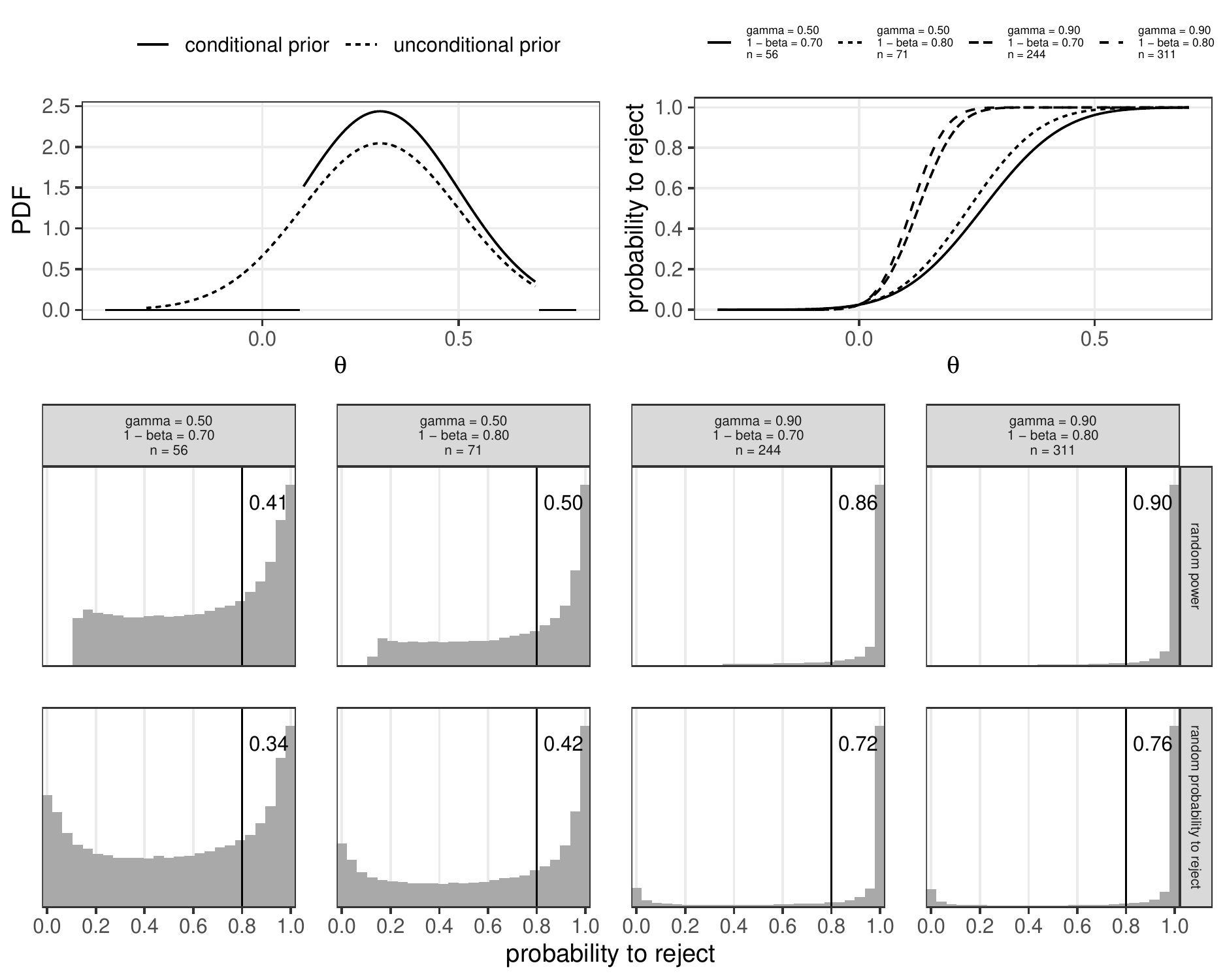}
    \caption{%
        Top: conditional and unconditional prior PDF for a truncated normal prior on $[-0.3, 0.7]$ with mean 0.3, standard deviation 0.2 and
        power curves corresponding to $n_{\gamma}^*$ for $1-\beta=0.8$ or $0.7$ and $\protect\mcid=0.1$;
        Bottom: histogram of the random power (conditional on relevant effect) and the random probability to reject (unconditional); vertical lines mark 80\% power and the number are the respective (conditional) probabilities to exceed a probability to reject of either 70\% or 80\%.
    }
    \label{fig:power-distribution-quantile-approach}
\end{figure}

To compare the results in the previous section with the prior quantile-based approach, we consider the intermediate example with prior mean 0.3 and prior standard deviation 0.2 again.
For this situation, the required sample sizes under $\gamma=0.5,0.9$ and target power $0.7$ or $0.8$, the corresponding curves of the rejection probability, and histograms of the distribution of the rejection probability are given in Figure~\ref{fig:power-distribution-quantile-approach}.

The required sample sizes depend heavily on the choice of $\gamma$.
The crucial difference between the quantile-based and the expected power approach is that for the prior quantile approach the exact distribution of power below the target value of 80\% is irrelevant; only the total mass of the distribution below this critical point matters.
This means that the sample size for the $\gamma=0.5$ cases are substantially lower than the corresponding sample size derived from an expected power constraint.
The flip side of this ignorance about \emph{the exact amount} by which the target power is undershot is that there is a relatively high chance of ending up with a severely underpowered study in these cases.
Increasing the certainty to exceed a power of 80\% or 70\% by setting $\gamma=0.9$, however, leads to substantially larger required sample sizes than under the expected power approach.

The example demonstrates the problems arising from having to specify both $\gamma$ and $1-\beta$.
While this allows more fine-grained control over the distribution of the (conditional) rejection probability, there seems to be no canonical choice for $\gamma$, which is critical in determining the required sample size.

\subsection{A clinical trial example}
\label{sec:example:clinical-trial}

To make things more tangible, consider the case of a clinical trial designed to demonstrate superiority of an intervention over a historical control group with respect to the endpoint of overall survival.
To stay within the framework of (approximately) normally distributed test statistics, we assume that effect sizes are given on a standardised log hazard scale, i.e., $\theta=0$ corresponds to no difference in overall survival and $\theta>0$ to superiority of the intervention group.
Assume that the prior for the treatment effect of the intervention is given by a truncated Normal distribution on $[\,-0.3, 0.7\,]$ with mean $0.2$ and standard deviation $0.2$ (pre-truncation).
The minimal clinically relevant difference is set to $\mcid=0.05$.
This setting corresponds to an \textit{a~priori} probability of a relevant effect of approximately $0.86$.

\begin{figure}
    \centering
    \includegraphics[width=\textwidth]{%
        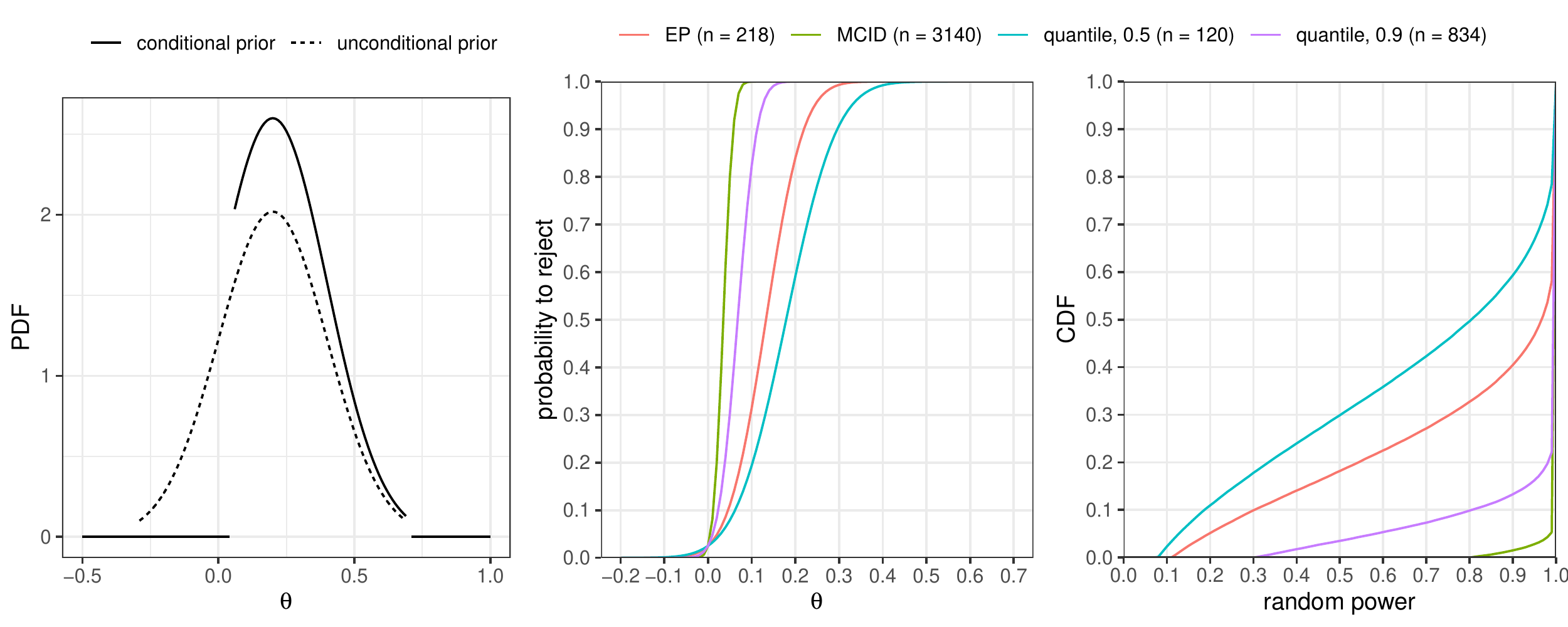}
    \caption{Left panel: prior PDF and conditional prior PDF (on $\Theta>\mcid = 0.05$); middle panel: probability to reject the null hypothesis as function of $\theta$ for the expected power design (EP), the design powered for $\mcid$ (MCID), the design based on power at the conditional prior median (quantile 0.5),
    and the design using the conditional 0.1-quantile, i.e. $\gamma=0.9$ (quantile 0.9); right panel: CDF of random power (probability to reject given $\Theta>\mcid = 0.05$) for the four different design choices.}
    \label{fig:example}
\end{figure}

Figure~\ref{fig:example} shows the (conditional) prior density, the curves of the rejection probability corresponding to the required sample sizes derived from constraints on a minimal probability to reject of $1-\beta=0.8$ at $\mcid$ (MCID), at ${\quant{0.5}[\,\Theta\geq\mcid\,]\approx0.26}$ (quantile, 0.5), at ${\quant{0.9}[\,\Theta\geq\mcid\,]\approx0.10}$ (quantile, 0.9), or a minimal expected power of $1-\beta=0.8$ (EP).

In this case the MCID criterion requires $n=3140$.
The quantile approach (with $\gamma=0.9$) already reduces this to $n=834$ while still maintaining an \textit{a~priori} chance of 90\% to exceed the target power of 80\%.
The quantile approach with $\gamma=0.5$ results in the lowest sample size of $n=120$ at the cost of only having a 50\% chance to exceed the target power of 80\%.
The EP approach is more liberal than the quantile approach ($\gamma=0.9$) with $n=218$ but still guarantees a chance of exceeding the target power of roughly 75\%.
A sample size based on $\PoS(n)\geq1-\beta=0.8$ cannot be derived in this example since the \textit{a~priori} probability of a relevant value is lower than $0.8$.
The large spread between the derived sample sizes shows how sensitive the
the required sample size is to the changes in the power constraint.
Clearly, the MCID approach is highly ineffective, as accepting a small chance to undershoot the target power with the quantile approach ($\gamma=0.9$) reduces the required sample size from $n=3140$ to roughly a quarter ($n=834$).
At the other extreme, constraining power only on the conditional prior median (quantile approach, $\gamma=0.5$) leads to a rather unattractive \textit{a~priori} distribution of the random power:
by definition, the probability to exceed a rejection probability of $0.8$ is still $0.5$ but the \textit{a~priori} chance of ending up with a severely underpowered study is non-negligible.

These considerations leave the trial-sponsor with essentially two options.
Either a range of scenarios for the quantile approach with values of $\gamma$ between $0.5$ and $0.9$ could be discussed in more detail and a decision on the exact value of $\gamma$ could be reached by considering the corresponding distributions of $\rpower{n}$, or the intermediate EP approach could be used.
We assume that the trial-sponsor accepts the implicit trade-off inherent to expected power and decides to base the sample size derivation on the EP approach.
The required sample size for an (expected) power of 80\% is then $n=218$.
Note that this still means that there is a roughly one-in-five \textit{a~priori} probability to end up in a situation with less than 50\% power (see Figure~\ref{fig:example}, CDF panel).

\begin{figure}
    \centering
    \includegraphics[width=\textwidth]{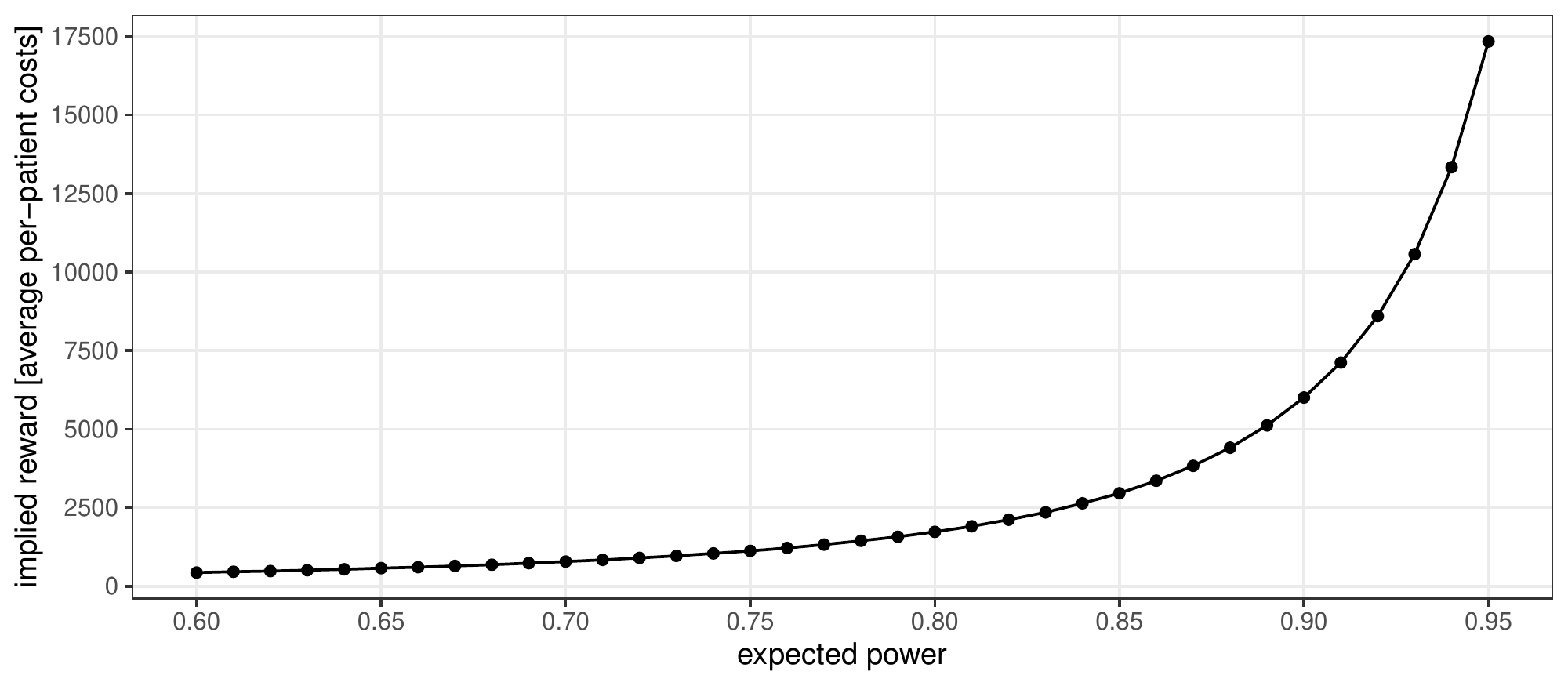}
    \caption{%
        Utility-maximising implied reward $\lambda$ for varying expected power levels in the situation discussed in Section~\ref{sec:example:clinical-trial}.
    }
    \label{fig:matched-reward}
\end{figure}

In a situation where $1-\beta=0.8$ is not set in stone, further insights might be gained by making the link to
utility maximisation explicit.
In a first step, we will assume that the sponsor has no way of quantising the reward parameter $\lambda$ directly.
One may then guide decision making by computing the values of $\lambda$ that lead to the same required sample size for a range of values of $\beta$.
Figure~\ref{fig:matched-reward} shows this `implied reward' as function of the minimal expected power constraint.
An expected power of $0.8$ is thus ideal in this situation if the expected reward upon successful (i.e., the effect is indeed relevant) rejection of the null hypothesis is approximately $1732$ times the average per-patient costs within the planned trial.
Using the curve depicted in Figure~\ref{fig:matched-reward}, a discussion with the trial sponsor about the plausibility of certain reward levels can be started.
Usually the average per-patient costs are well-known in advance, so that the scale can even be transformed to monetary units, e.g., \$US.
Assume to this end, that for the particular study at hand, the expected average per-patient costs are $30\,000$ \$US.
Then, the sample size corresponding to an expected power of $0.8$ is maximising utility if the expected reward is $30\,000\cdot1732=51.96\cdot 10^6$ \$US.
The utility-maximising reward for an expected power of $0.9$ would be approximately $6006$, i.e., $180.18\cdot 10^6$ \$US.
Even without committing to a fixed value of $\lambda$, these considerations can be used to guide the decision as to which of the `standard' power levels (0.8 or 0.9) might be more appropriate in the situation at hand.

Of course, one might also directly optimise utility if the reward upon successful rejection of the null hypothesis can be specified.
To that end, assume that a reward of $100\cdot 10^6$ \$US is expected.
Under the same assumption about average per-patient costs, this translates to $\lambda\approx3333$.
The utility-maximising sample size is then $n=329$ and the corresponding utility-maximising expected power is $0.86$.

\section{Discussion}

The concept of `hybrid' sample size derivations based on
Bayesian priors for planning and the design's frequentist error rate properties is well-established in the literature on clinical trial design.
Nevertheless, the substantial variation in the terminology used and small differences in the exact definition of the terms used can be confusing.
We have tried our best to formulate a consistent naming scheme,
to be explicit about the exact definitions,
highlight connections between the different quantities
(see Figure~\ref{fig:diagram}),
and to relate back to previous authors (see Table~\ref{tab:literature-overview}).
Any naming scheme necessarily has a subjective element to it
and ours is by no means exempt from this problem (see also \url{https://xkcd.com/927/}).
We do hope, however, that our review encourages a clearer separation
between terminology for joint probabilities (avoiding the use of the word `power') and for probabilities that condition on the presence of an effect (`power' seems more appropriate here), as well as a more transparent distinction between arguments based on \textit{a~priori} likelihood of effects and their relevance.
We also strongly believe that an explicit definition (in formulae)
of any quantities used should be given when discussing
the subject.
Merely referring to terms like `expected power' or
`probability of success' are too ambiguous given their inconsistent use in the literature.

Often, the main argument for a `hybrid' approach to sample size derivation is the fact that the uncertainty about the true underlying effect can be incorporated in the planning of a design.
This is certainly a major advantage but it is equally important that
the `hybrid' approach allows a very natural distinction between
arguments relating to the (relative) \textit{a~priori} likelihood of different parameter values (encoded in the prior density)
and relevance arguments (encoded in the choice of $\mcid$).
The fact that these two components can be represented naturally within the `hybrid' approach has the potential to make sample size derivation much more transparent.

The `hybrid' quantity considered most commonly in the literature is the marginal probability to reject $\mathcal{H}_0$.
Often, it is not clear whether the authors are aware of the fact that this quantity includes the error of rejecting the null hypothesis incorrectly, i.e. when $\theta<\mcid$. 
In many practical situations this problem is numerically negligible and $\PoS'(n)\approx\PoS(n)$, i.e.,~the marginal probability to reject the null hypothesis is approximately the same
as the joint probability of a non-null effect and the rejection of the null hypothesis.
If, however, the definition of `success' also takes into account a non-trivial relevance threshold $\mcid>\theta_0$,
the distinction becomes more important in practice.
Given the great emphasis on strict type~I error rate control in the clinical trials community it seems at least strange to implicitly consider type~I errors as `successful' trial outcomes.
Beyond these principled considerations,
a practical advantage of $\PoS(n)$ over $\PoS'(n)$ is the
direct and simple connection to $\EP(n) = \PoS(n) / \Pr[\,\Theta\geq\mcid\,]$.
While $\EP(n)$ is independent of the \textit{a~priori} probability
of a relevant effect and only depends on the relative \textit{a~priori} likelihood of different effects through the
conditional prior, $\PoS(n)$ does directly depend on $\Pr[\,\Theta\geq\mcid\,]$.
Although \citet{spiegelhalter-2004} see this as a disadvantage of $\EP(n)$, it is actually a necessary property to use it for sample size derivation without re-calibrating the conventional values for $1-\beta$ (see also \citealp{brown1987projection}).
If one tried to derive a sample size such that $\PoS(n)=0.8$ this would be impossible for situations with $\Pr[\,\Theta\geq\mcid\,]<0.8$.
In a situation where $\Pr[\,\Theta\geq\mcid\,]$ exceeds 0.8 only slightly, the expected power ($\EP(n)$) would have to be close to 1 to compensate for the \textit{a~priori} probability of a relevant effect.
In essence, one would thus increase the sample size in situations where the efficacy of the new treatment is still uncertain.
This would put more study participants at risk just to make sure that the treatment effect is detected almost certainly if it is indeed present.
The use of $\PoS(n)$ for sample size derivation thus only makes sense in a setting where the threshold $1 - \beta$ is adapted to the \textit{a~priori} probability of a relevant effect.
The simplest way to do so is by using $\PoS(n) \geq \Pr[\,\Theta\geq\mcid\,]\,(1 - \beta)$ which is, however, entirely equivalent to $\EP(n) \geq 1 - \beta$.
Another option to derive situation-specific thresholds is via utility maximisation, and $\PoS(n)$ is a key term in the simple expected utility function proposed in Section~\ref{sec:utility}.
Ultimately, $\PoS(n)$ and $\EP(n)$ can be used interchangeably once the prior distribution is fixed as long as the respective multiplicative factor is taken into account.
The main advantage of $\PoS(n)$ is that it is an unconditional probability
which might be easier to interpret by practitioners, while $\EP(n)$ can be readily used in conjunction with an already established power threshold in a research field.

A slightly different concept to sample size derivation via expected power is what we call the `quantile approach'.
This approach uses a different functional of the probability to reject the null hypothesis given a relevant effect.
Instead of the mean, we propose to use a $\gamma$ quantile of this distribution.
Compared to expected power, this allows direct control of
the left-tail of the \textit{a~priori} distribution of the probability to reject the null hypothesis given a relevant effect.
This can be desirable since a sample size derived via a threshold for expected power might still lead to a substantial chance of ending up with an underpowered study.
This can be avoided with the quantile approach and a higher value for $\gamma$ (see Figure~\ref{fig:power-distribution-quantile-approach}).
The quantile approach is also relatively easy to implement in practice, since it is just a Bayesian justification for powering on a point alternative.
This flexibility comes at the price of having to specify an additional parameter, $\gamma$ (the acceptable risk of ending up with an underpowered study).
Theoretically, both expected power and the prior quantile approach are perfectly viable to determine a sample size.
Whichever approach is preferred, it is certainly advisable to
not only plot the corresponding power curves but also the resulting distribution of $\rpower{n}$ (see Figure~\ref{fig:power-distribution}).
In essence, the problem of defining a `hybrid' power constraint boils down to finding a summary functional of the power curve that reflects the planning objectives.
Ideally, one would like to control the \textit{a~priori} distribution of $\rpower{n}$ such that it is sharply peaked around a certain target value avoiding both over- and underpowered studies.
Yet, controlling both location (e.g.,~mean) and spread (e.g.,~standard deviation) of the distribution of $\rpower{n}$ is
impossible.
A second constraint on the standard deviation of $\rpower{n}$
in addition to the mean constraint (expected power)
would led to an over-determined problem since there is only one free parameter,~$n$.
To increase expected power, the sample size must be increased.
The standard deviation of $\rpower{n}$, however, decreases as the sample size is lowered since this flattens the power curve of the resulting test (the standard deviation would be 0 if the power curve was constant).
Both conflicting objectives (high expected power, low standard deviation of power) are thus not fulfillable at the same time.

Finally, it should be stressed again that the key frequentist property of strict type~I error rate control of the designs are not affected by the fact that the arguments for calculating a required sample size are Bayesian.
In fact, at no point, the Bayes theorem is invoked (i.e.~the posterior distribution of the effect size is not required).
The Bayesian perspective is merely a principled and insightful way of specifying a weight function (prior density) that can then be used to guide the choice of the power level of the design, or as \citet{brown1987projection} put it: ``This proposed use of Bayesian methods should not be criticised by frequentists in that these methods do not replace any current statistical techniques, but instead offer additional guidance where current practice is mute''.

\subsection*{Supplemental Materials}

The code required to reproduce the figures is available at \url{https://github.com/kkmann/sample-size-calculation-under-uncertainty}.
A permanent backup of the exact version of the repository used for this manuscript is available under the digital object identifier
\href{https://doi.org/10.5281/zenodo.3899943}{10.5281/zenodo.3899943} (release 0.2.1).
An interactive version of the repository at the time of publication is
hosted at \url{https://mybinder.org/v2/gh/kkmann/sample-size-calculation-under-uncertainty/0.2.1?urlpath=lab/tree/notebooks/figures-for-manuscript.ipynb} using Binder~\citep{jupyter-2018}.
A simple shiny app implementing the sample size calculation procedures is
available at \url{https://mybinder.org/v2/gh/kkmann/sample-size-calculation-under-uncertainty/0.2.1?urlpath=shiny/apps/sample-size-calculation-under-uncertainty/}.

\subsection*{Funding}

DSR was funded by the Biometrika Trust and the Medical Research Council (MC\_UU\_00002/6).

\subsection*{Conflicts of interest}

None to declare.

\bibliography{references.bib}

\end{document}

%% file: diagram.tex
\begin{tikzpicture}[->,>=stealth']
\footnotesize
    \node[state,
        text width=4.5cm,
        yshift=0cm,
        anchor=center,
        align=left,
        inner sep=.75em
    ] (power) {
        \begin{tabular}{rl}
            \multicolumn{2}{l}{\textbf{Probability to reject $\mathcal{H}_0$}} \\[1em]
            \textit{Sym.}: & 
                $\Pr_\theta[\,Z_n > \criticalvalue\,]$ \\[.33em]
            \textit{Def.}: & 
                $\Phi\big(\theta_n - \criticalvalue\big)$ \\[.33em]
            \textit{Int.}: & 
                \parbox[t]{3.25cm}{
                real number; 
                probability to reject the null hypothesis \emph{given a~fixed~value~of~$\theta$}
                }
        \end{tabular}
    };
    
    \node[state,
        text width=8.25cm, 	
        yshift=9cm, 		
        right of=power, 	
        node distance=7.5cm, 	
        anchor=center,
        align=left,
        inner sep=.75em
    ] (rprreject) {
        \begin{tabular}{rl}
            \multicolumn{2}{l}{\textbf{Random probability to reject $\mathcal{H}_0$}} \\[1em]
            \textit{Sym.}: & $\rpr(n)$ \\[.33em]
            \textit{Def.}: & $\Pr_{\Theta}[\,Z_n > \criticalvalue\,] = \Phi\Big(\sqrt{n}/ \sigma\,(\Theta - \theta_0) - \criticalvalue\Big)$ \\[.33em]
            \textit{Int.}: & \parbox[t]{6.5cm}{random variable; realisations correspond to the probability to reject the null hypothesis for $\Theta=\theta$}
        \end{tabular}
    };
    
    \node[state,
        text width=8.25cm, 
        yshift=5cm, 		
        right of=power, 
        node distance=7.5cm, 
        anchor=center,
        align=left,
        inner sep=.75em
    ] (rpower) {
        \begin{tabular}{rl}
            \multicolumn{2}{l}{\textbf{Random power}} \\[1em]
            \textit{Sym.}: & $\rpower{n}$ \\[.33em]
            \textit{Def.}: & $\rpr(n) \mid \Theta \geq \mcid$ \\[.33em]
            \textit{Int.}: & 
                \parbox[t]{6.5cm}{random variable; realisations correspond to the probability to reject the null hypothesis for $\Theta=\theta$ \emph{given a relevant effect}}
        \end{tabular}
    };
    
    \node[state,
        text width=8.25cm, 	
        yshift=0.5cm, 		
        right of=power, 	
        node distance=7.5cm, 	
        anchor=center,
        align=left,
        inner sep=.75em
    ] (epower) {
        \begin{tabular}{rl}
            \multicolumn{2}{l}{\textbf{Expected power}} \\[1em]
            \textit{Sym.}: & $\EP(n)$ \\[.33em]
            \textit{Def.}: & \parbox[t]{3cm}{
                \vspace*{-1.8em}
                \begin{align}
                & \Pr[\,Z_n > z_{1-\alpha}\mid \Theta \geq\mcid\,]= \E[\,\rpower{n}\,]\nonumber\\ 
                &=\int_{\mcid}^\infty \Pr_\theta[\,Z_n > \criticalvalue\,]\ \varphi(\theta\,|\,\Theta\geq \mcid)\, \operatorname{d} \theta \nonumber
            \end{align}} \\[.33em]
            \textit{Int.}: & 
                \parbox[t]{6.5cm}{real number; average probability to reject the null weighted with prior density \emph{conditional on $\Theta\geq\mcid$}}
        \end{tabular}
    };
    
    \node[state,
        text width=8.25cm, 	
        yshift=-4.5cm, 		
        right of=power, 	
        node distance=7.5cm, 	
        anchor=center,
        align=left,
        inner sep=.75em
    ] (pos) {
        \begin{tabular}{rl}
            \multicolumn{2}{l}{\textbf{Probability of success}} \\[1em]
            \textit{Sym.}: & $\PoS(n)$ \\[.33em]
            \textit{Def.}: & \parbox[t]{3cm}{
                \vspace*{-1.75em}
                \begin{align}
                    &\Pr[\,Z_n > z_{1-\alpha}, \Theta \geq\mcid\,] \nonumber \\ 
                    &= \int_{\mcid}^\infty \Pr_\theta[\,Z_n > \criticalvalue\,]\ \varphi(\theta)\, \operatorname{d}\theta \nonumber
                \end{align}
            } \\[.33em]
            \textit{Int.}: & 
                \parbox[t]{6.5cm}{real number; joint probability to reject the null and have a relevant effect; average probability to reject on $\Theta\geq\mcid$ weighted with \emph{unconditional} prior density}
        \end{tabular}
    };
    
    \node[state,
        text width=8.25cm, 	
        yshift=-9.5cm, 		
        right of=power, 	
        node distance=7.5cm, 	
        anchor=center,
        align=left,
        inner sep=.75em
    ] (posprime) {
        \begin{tabular}{rl}
            \multicolumn{2}{l}{\textbf{Marginal probability to reject $\mathcal{H}_0$}} \\[1em]
            \textit{Sym.}: & $\PoS'(n)$ \\[.33em]
            \textit{Def.}: & \parbox[t]{3cm}{
                \vspace*{-1.75em}
                \begin{align}
                    &\Pr_{\varphi(\cdot)}[\,Z_n > z_{1-\alpha}\,] = \E[\,\rpr(n)\,] \nonumber \\
                    &=  \int_{-\infty}^\infty \Pr_\theta[\,Z_n > \criticalvalue\,]\ \varphi(\theta)\ \operatorname{d}\theta \nonumber
                \end{align}
            } \\[.33em]
            \textit{Int.}: & 
                \parbox[t]{6.5cm}{real number; marginal probability to reject the null irrespective of underlying effect}
        \end{tabular}
    };
  
    \path 
    (power) edge[bend left=20] 
        node[anchor=south,above,xshift=-1.25cm]{%
            \parbox{2cm}{\tiny
                replace fixed parameter $\theta$ with random variable ${\Theta\sim\varphi(\cdot)}$}
            }
        (rprreject.west)
    (rprreject) edge
        node[anchor=west,xshift=.25em]{%
            \parbox{2.5cm}{\tiny
                condition on ${\Theta\geq\mcid}$}
            }
        (rpower)
    ([xshift=0.5cm]power.north) edge[bend left=12.5]
        node[anchor=south,above,yshift=0.15cm,xshift=-.5em]{%
            \parbox{2.5cm}{\tiny
                integrate with respect to prior conditional on relevant effect, ${\varphi(\cdot\mid\Theta\geq\mcid)}$}
            }
        (epower)
    (power) edge[bend right=30]
        node[anchor=east,above,xshift=2em, yshift=.33cm]{%
            \parbox{1.66cm}{\tiny
                integrate over $\theta \geq \mcid$ with unconditional prior $\varphi(\cdot)$}
            }
        (pos.west)
    (power) edge[bend right=20] 
        node[xshift=-1.15cm]{%
            \parbox{1.55cm}{\tiny
                integrate over entire parameter range with unconditional prior $\varphi(\cdot)$}
            }
        (posprime.west)
    (rpower) edge
        node[anchor=west,xshift=.25em]{%
            \parbox{2.5cm}{\tiny
                form expected value}
            }
        (epower)
    (epower) edge
        node[anchor=west,xshift=.5em]{%
            \parbox{3cm}{\tiny
                $\cdot \Pr_{}[\,Z_n > \criticalvalue\,]$}
            }
        (pos)
    (pos) edge
        node[anchor=west,xshift=.5em]{%
            \parbox{3.5cm}{\tiny
                add expected type one error rate and probability of rejection under irrelevant values}
            }
        (posprime)
    (rprreject.east) edge[bend left=10]
        node[anchor=east,below,xshift=1.5em,yshift=-2.25cm]{%
            \parbox{1cm}{\tiny
                form expected value}
            }
        (epower.east)
    (rprreject.east) edge[bend left=15]
        node[anchor=east,below,xshift=-1.5em,yshift=-1cm]{%
            \parbox{1cm}{\tiny
                form expected value}
            }
        (posprime.east);

\end{tikzpicture}

%% file: literature-review-table.tex
\textbf{Concept} & \textbf{References} & \textbf{Notes} \\
\hline
\endhead
\textbf{Marginal probability to reject $\mathcal{H}_0$} & 
    \cite{crook-1982} & 
        Termed `strength'; application in multinomial contingency tables. \\
    &
    \cite{spiegelhalter1986} &
        Only implicitly mentioned; discussing close relation to $\PoS(n)$, termed `expected/average power' in \citet{spiegelhalter-2004}.\\
    &
    \cite{gillett-1994} & 
    Termed `average power`; focus on replication. \\
&
\cite{o2001bayesian} &
    Termed `assurance' or `expected power'; different from our notion of expected power which is conditional on a relevant effect, see also \citep{ohagan-2005}. \\
&
\cite{chuang-2006} &
    Termed `average probability of success';
    discusses other definitions of `success' based on additional criteria for the observed point estimates;
    discusses how basing the sample size on relevance arguments alone is theoretically correct but ineffective if evidence for larger effect sizes is available, see also \citet{chuang2011}.\\
&
\cite{grouin-2007} & 
    Termed `predictive power' and `predictive probability to reject $\mathcal{H}_0$'; review of regulatory aspects, discussion of interval-based sample size calculation, and utility considerations. \\
&
\cite{daimon-2008} & 
    Termed `hybrid Neyman–Pearson–Bayesian (hNPB) probability`; application in non-inferiority setting. \\
&
\cite{shao-2008} & 
    Termed `adjusted power'; review of regulatory aspects, discussion of interval-based sample size calculation, and utility considerations. \\
&
\cite{liu-2010} & 
        Termed `extended Bayesian expected power 1'; extended by treating variance as unknown, also consider $\PoS(n)$ and $\EP(n)$. \\
&
\cite{lan-2012} & 
    Termed `average power'; discusses upper limit of `average power` depending on prior choice and suggest truncated priors which would be very close to conditioning on a relevant effect. \\
&
\cite{carroll-2013} &
    Termed `assurance' and `probability of success'
    ($\operatorname{PoS}$);
    discusses other definitions of success but all definitions are also exclusively based on \emph{observed}
    quantities (minimum threshold on point estimate), 
    see also \citet{chuang-2006}. \\
&
\cite{brutti-2014} & 
    Termed `predictive frequentist power'; also discusses sample size derivation based on Bayesian decision criteria. \\
&
\cite{ren-2014} & 
    Termed `assurance'; discusses ideas of \citet{ohagan-2005} in time-to-event setting. \\
&
\cite{hu-2014} & 
    Termed `probability of success'; 
    considers priors on mean and standard deviation; discuss upper limit on probability of success in the more complex two-parameter situation. \\
&
\cite{ibrahim-2015} & 
    Termed `average probability of success'; 
    discussed in context of historical data integration. \\
& 
\cite{walley-2015} & 
    Termed `assurance' or `probability of success'; 
    extension to multi-parameter situations. \\
& 
\cite{ciarleglio-2015} & 
        Termed `expected power'; also consider $\EP(n)$ and $\PoS(n)$, very similar settings considered in \citet{ciarleglio-2016,ciarleglio-2017}.\\
&
\cite{rufibach_15} & 
    Termed `assurance' or `probability of success'; 
    in-depth discussion of the distribution of the probability to reject the null hypothesis. \\
& 
\cite{saint-hilary-2018} & 
    Termed `predictive probability of success';
    consider both `statistical success' ($p$-value $\leq\alpha$) and `clinical relevance' (\emph{observed}~effect above relevance threshold), see also \citet{saint-hilary-2019}. \\
&
\cite{chen-2017} &
    Termed `assurance' and `expected power'; discusses conditional nature of the (frequentist) probability to reject the null hypothesis from a Bayesian perspective. \\
& 
\cite{jiang-2011,kirby-2012,zhang-2013,wang-2015,gotte-2017} &
    Termed `probability of statistical success', `probability of success', `assurance', `predictive power'; discusses extensions to multiple studies or entire drug development programs.
    \\
&
\cite{ambrosius-2012,wang-2013,wang-2015b,crisp-2018,chen-2018} & 
    Termed `assurance', `probability of success', `probability of study success'; practical applications in various settings. \\
\textbf{Probability of success} &
    \cite{spiegelhalter1986} &
        Only implicitly mentioned, termed `prior adjusted power' in \citet{spiegelhalter-2004}; discusses close relation to marginal probability to reject $\mathcal{H}_0$ (suggesting the latter as practical approximation).\\
    &
    \cite{brown1987projection} & 
        Termed `expected power'; also discusses `conditional expected power' which corresponds to our definition of $\EP(n)$. \\
    &
    \cite{shao-2008} & 
        Termed `adjusted power`; application of the ideas of \citet{spiegelhalter-2004} to binary setting, define probability of success but approximate it with the marginal probability to reject $\mathcal{H}_0$. \\
    &
    \cite{liu-2010} & 
        Termed `extended Bayesian expected power 2'; extended by treating variance as unknown, also considers $\PoS'(n)$ and $\EP(n)$. \\
    &
    \cite{ciarleglio-2015} & 
        Termed `prior-adjusted power'; also considers $\EP(n)$ and $\PoS'(n)$, very similar settings considered in \citet{ciarleglio-2016,ciarleglio-2017}.\\
\textbf{Expected power} & 
    \cite{brown1987projection} & 
        Termed `conditional expected power'; also discusses unconditional expected power which corresponds to our definition of $\PoS(n)$. \\
    &
    \cite{spiegelhalter-2004} &
        Not named; referencing \cite{brown1987projection}.\\
    &
    \cite{liu-2010} & 
        Termed `extended Bayesian expected power 3'; extended by treating variance as unknown, also consider $\PoS(n)$ and $\PoS(n)$. \\
    &
    \cite{ciarleglio-2015} & 
        Termed `conditional expected power'; also considers $\PoS(n)$ and $\PoS'(n)$, very similar settings considered in \citet{ciarleglio-2016,ciarleglio-2017}.\\
\hline